\documentclass[preprint]{aastex}
\usepackage{emulateapj5}

\input psfig.tex

\def\hst{{\it HST}}

\def\etal{\emph{et al.}\ }

\def\kms{km s$^{-1}$}
\def\apj{ApJ}
\def\aj{AJ}
\def\mnras{MNRAS}
\def\apjs{ApJS}

\def\aa{{A\&A}}
\def\aas{{ A\&AS}}
\def\aj{{AJ}}
\def\al{$\alpha$}
\def\bet{$\beta$}
\def\amin{$^\prime$}
\def\annrev{{ARA\&A}}
\def\apj{{ApJ}}
\def\apjs{{ApJS}}
\def\asec{$^{\prime\prime}$}

\def\deg{$^{\circ}$}
\def\ddeg{{\rlap.}$^{\circ}$}

\def\e#1{$\times$10$^{#1}$}
\def\etal{{et al. }}
\def\flamb{erg s$^{-1}$ cm$^{-2}$ \AA$^{-1}$}

\def\lamb{$\lambda$}
\def\lum{erg s$^{-1}$}
\def\micron{{$\mu$m}}
\def\mnras{{MNRAS}}
\def\nat{{Nature}}
\def\pasp{{PASP}}

\def\percm2{cm$^{-2}$}

\def\lax    {${_<\atop^{\sim}}$ }
\def\gax    {${_>\atop^{\sim}}$ }

\def\hii{\ion{H}{2}}
\def\oiii{[\ion{O}{3}]}

\slugcomment{To appear in The Astrophysical Journal.}
\shorttitle{{LUMINOSITIES OF SEYFERT NUCLEI}}
\shortauthors{HO \& PENG}

\begin{document}

\title{Nuclear Luminosities and Radio Loudness of Seyfert Nuclei}

\author{Luis C. Ho}
\affil{The Observatories of the Carnegie Institution of Washington, \\
813 Santa Barbara St., Pasadena, CA 91101}

\and 

\author{Chien Y. Peng}
\affil{Steward Observatory, Univ. of Arizona, Tucson, AZ 85721}

\begin{abstract}
Historically, Seyfert nuclei have been considered to be radio-quiet active 
galactic nuclei (AGNs).  We question this widely held assumption by showing 
that the distribution of the radio-to-optical luminosity ratio, 
$R\,\equiv\,L_{\nu}({\rm 6~cm})/L_{\nu}(B)$, when properly measured 
for the {\it nuclear}\ component, places the majority of type~1 Seyfert nuclei 
in the category of radio-loud AGNs, defined here as objects with $R\,\geq\,10$.
This result is further strengthened by strong correlations found between
radio power and optical continuum and emission-line luminosities, as has been 
established previously for more powerful AGNs.  We also present a new 
calibration of the relation between optical continuum and Balmer emission-line 
luminosities valid in the regime of low-luminosity AGNs.

\end{abstract}


\section{Introduction}

Active galactic nuclei (AGNs) nearly universally emit synchrotron radiation at 
radio wavelengths.  The character of the radio emission shows tremendous 
diversity, ranging from relatively simple structures such as parsec-scale 
central cores, to well collimated jets and lobes extending over hundreds of 
kpc.  Historically, one of the most notable attributes of AGNs is the apparent 
bimodal distribution of their strength of radio emission (Strittmatter et al. 
1980; Sramek \& Weedman 1980).  Optically selected AGNs seem to fall into one 
of two categories --- ``radio loud'' or ``radio quiet'' --- the dividing line 
generally set by the relative monochromatic luminosities at radio and optical 
wavelengths. A widely adopted convention (or some closely related variant) is 
$R \,\equiv\,L_{\nu}({\rm 6~cm})/L_{\nu}(B)$, with the boundary between the 
two radio classes set at $R$ = 10 (Visnovsky et al. 1992; Stocke et 
al. 1992; Kellermann et al. 1994).  Miller, Peacock, \& Mead (1990) advocate 
an alternative criterion based on the radio luminosity alone, a limit set to 
$P_{\rm 6cm}\,\approx\,10^{25}$ W~Hz$^{-1}$ sr$^{-1}$.  The majority of 
optically selected quasars are radio quiet, with only $\sim$10\%--20\% 
qualifying as radio loud, based on either of the above criteria (e.g., 
Kellermann et al. 1989; Visnovsky et al. 1992; Hooper et al. 1995).   Some 
recent studies using radio selected samples of quasars have questioned the 
bimodality in the distribution of $R$ (Wadadekar \& Kembhavi 1999; White 
et al. 2000).  

The actual shape of the $R$ distribution notwithstanding, the fraction of 
radio-loud AGNs tends to increase for the optically most luminous objects 
(e.g., Miller et al. 1990).  At low redshift, where high-resolution optical 
and near-infrared images are now readily available, radio-loud quasars 
preferentially inhabit luminous, massive, early-type galaxies, whereas the 
hosts of radio-quiet quasars span a wider range of morphological types (e.g., 
McLure et al. 1999; Hamilton, Casertano, \& Turnshek 2001).  Radio galaxies 
themselves have long been known to be predominantly giant ellipticals 
(e.g., Matthews, Morgan, \& Schmidt 1964; Zirbel 1996; de~Vries et al. 2000).  
Within this backdrop, Seyfert galaxies, which are generally considered to be 
the low-luminosity extension of the quasar phenomenon, traditionally have been 
regarded as radio-quiet objects. This perception perhaps has been reinforced 
further by the apparent correlation between radio loudness and host galaxy 
morphology, since most Seyfert nuclei reside in disk (spiral and lenticular) 
systems (e.g., Adams 1977; Yee 1983; MacKenty 1990; Ho, Filippenko, \& Sargent 
1997b). By virtue of their proximity, a high percentage of Seyfert galaxies 
are detectable as radio sources (e.g., de Bruyn \& Wilson 1978; Ulvestad \& 
Wilson 1984a, 1984b, 1989; Edelson 1987; Rush, Malkan, \& Edelson 1996; Ho \& 
Ulvestad 2001).  Nonetheless, their modest radio luminosities ($P_{\rm 6cm}\, 
\approx\,10^{19}-10^{22}$ W Hz$^{-1}$) and their accompanying small 
radio-to-optical luminosity ratios have cemented the notion that Seyferts are 
radio-quiet objects.  In phenomenological classifications of AGNs, type~1 and 
type~2 Seyferts often are viewed as the radio-quiet analogs of broad-line and 
narrow-line radio galaxies, respectively (e.g., Osterbrock 1984; Lawrence 
1987; Blandford 1990; Peterson 1997; Krolik 1998).  Laor (2000) explicitly 
argues that the bulges of spiral galaxies cannot produce radio-loud AGNs.

Extending the $R$ parameterization to Seyferts, however, is not 
straightforward.  In the case of quasars, the active {\it nucleus}\ 
significantly outshines the underlying galaxy at optical, and often also 
at radio, wavelengths.  Consequently, $R$, computed from the integrated 
emission, faithfully represents a quantity germane to the AGN.  The situation 
for Seyferts is more complicated.  Once an object is recognized as a Seyfert 
{\it galaxy}, the underlying host obviously contributes appreciably to the 
integrated optical light of the system.  Most Seyfert nuclei are found in 
bulge-dominated, relatively luminous ($\sim L^*$) galaxies; typically, 
$M_{B_T}\,\approx$ $-21\pm 1$ mag (e.g., Ho et al. 1997b)\footnote{Throughout 
this paper, we adopt a Hubble constant of $H_0$ = 75 \kms\ Mpc$^{-1}$ and a
deceleration parameter of $q_0$ = 0.5.}.  The optical luminosities of the 
nuclei, on the other hand, span an enormous range.  Given the tremendous range 
of brightness contrast between the nucleus and the bulge, in general the 
optical luminosity of Seyfert nuclei can be characterized properly only with
measurements which sufficiently resolve the nuclear region.   A similar caveat 
applies to the radio data. The disks of spiral galaxies emit synchrotron radio 
emission at a level which easily rivals the nuclear emission.  Normal 
early-type spirals, for instance, have typical 20~cm powers of 
10$^{20}$--10$^{22}$ W~Hz$^{-1}$ (Hummel 1981; Condon 1987), within the range 
emitted by the nucleus.  In general, therefore, the AGN component of the 
radio signal in Seyferts can be isolated reliably only using high-resolution, 
interferometric data.

The primary goal of this paper is to reexamine the conventional claim that 
Seyfert nuclei, and by implication their spiral host galaxies, are mainly 
radio quiet.   We use high-resolution optical and radio continuum measurements 
to show that the nuclear values of $R$ in the majority of Seyferts place them 
in the category of radio-loud objects.  Our data also furnish insights into 
the physical relationship between radio power and optical continuum and 
emission-line luminosities.  We suggest that the fueling of radio jets may be 
closely linked with the disk accretion rate.  Lastly, we present a new 
calibration of the relation between optical continuum and Balmer emission-line 
luminosities for type~1 AGNs.

\section{The Sample}

For the purposes of this study, we choose only objects classified as 
Seyfert~1 nuclei, which are believed to be relatively unobscured sources and 
hence provide the most direct comparison with quasars. We consider a sample 
drawn from two optical spectroscopic surveys of nearby galaxies.  Ho, 
Filippenko, \& Sargent (1995) surveyed a nearly complete sample of 486 bright 
($B_T\,\leq$ 12.5 mag), northern ($\delta\,>$ 0\deg) galaxies using the 
Palomar 5-m telescope, from which they derived a sensitive catalog of 
emission-line nuclei, including a comprehensive list of nearby AGNs (Ho et al. 
1997a, 1997b, 1997c).  The Palomar survey contains 49 Seyfert 
galaxies\footnote{Ho et al. (1997a) list a total of 52 Seyferts (22 type~1, 30 
type~2), but three of these have $\delta\,<$ 0\deg\ and so formally do not 
belong to the complete sample.  We include intermediate-type objects 
(Osterbrock 1981) in the type~1 class.}, 21 of type~1 and 28 of type~2, most 
of which have low luminosities (median $L_{{\rm H}\beta}$ = 5\e{39} \lum) and 
are nearby (median $D$ = 20 Mpc).   As discussed by Ho et al. (1997b) and Ho 
\& Ulvestad (2001), the Palomar Seyfert sample is the most complete and least 
biased available.  One drawback of this sample, however, is that it contains 
relatively few objects on the upper end of the luminosity function.  

We therefore supplement the Palomar sample with Seyferts derived 
from the CfA redshift survey (Huchra et al. 1983).  Spectra were acquired 
for 2399 galaxies with Zwicky magnitudes $\leq$ 14.5 that lie within 
$\delta\,\geq$ 0\deg\ and $b_{\rm II}\,\geq$ 40\deg\ or $\delta\,\geq$ 
--2\ddeg5 and $b_{\rm II}\,\leq$ --30\deg.  Huchra \& Burg (1992) compiled a 
list of 49 Seyfert galaxies, the spectroscopic classifications of which were 
subsequently refined by Osterbrock \& Martel (1993).  According to the 
classifications of Osterbrock \& Martel, the CfA sample contains 
33 Seyfert~1s and 15 Seyfert~2s; one object (Mrk 789) was judged to be 
a starburst galaxy.  The median distance of the CfA Seyferts is $\sim$80 Mpc.
Ho \& Ulvestad (2001; see \S\ A.2) argue that the CfA Seyfert sample 
suffers from complex selection effects.  In the case of Seyfert~1s,
the primary effect is a bias against faint, intrinsically weak nuclei.  
As an illustration of the level of incompleteness, we note that 19 out of 
the 21 Seyfert~1s from the Palomar sample formally fall within the selection
boundaries of the CfA survey, but only nine (43\%) were recognized as 
Seyfert~1s by Huchra \& Burg; all nine objects have very prominent 
broad emission lines.   Thus, the samples of Seyfert~1 nuclei from the 
Palomar and CfA surveys complement one another, although the combined sample 
in itself cannot be regarded as complete.  Fortunately, this limitation 
does not affect the major results of this study, but we caution against other 
applications of the data where sample completeness matters (e.g., derivation 
of luminosity functions).  Our final sample, summarized in Table 1, consists 
of 45 objects: 21 from the Palomar survey, 33 from the CfA survey, and nine 
common to both.

\vskip 0.3cm
 
\begin{figure*}[t]
\centerline{\psfig{file=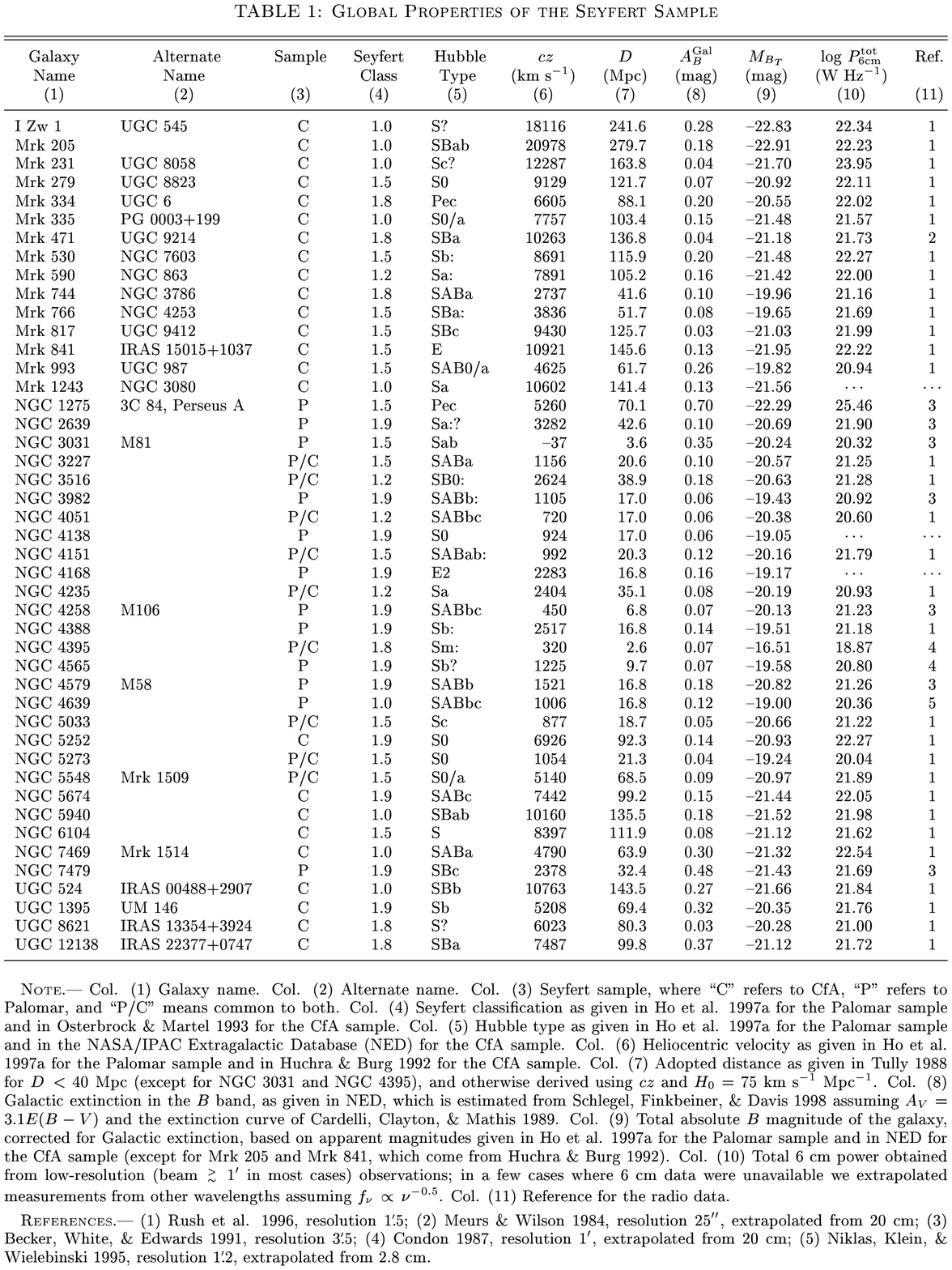,width=18.5cm,angle=0}}
\end{figure*}
 

 
\begin{figure*}[t]
\centerline{\psfig{file=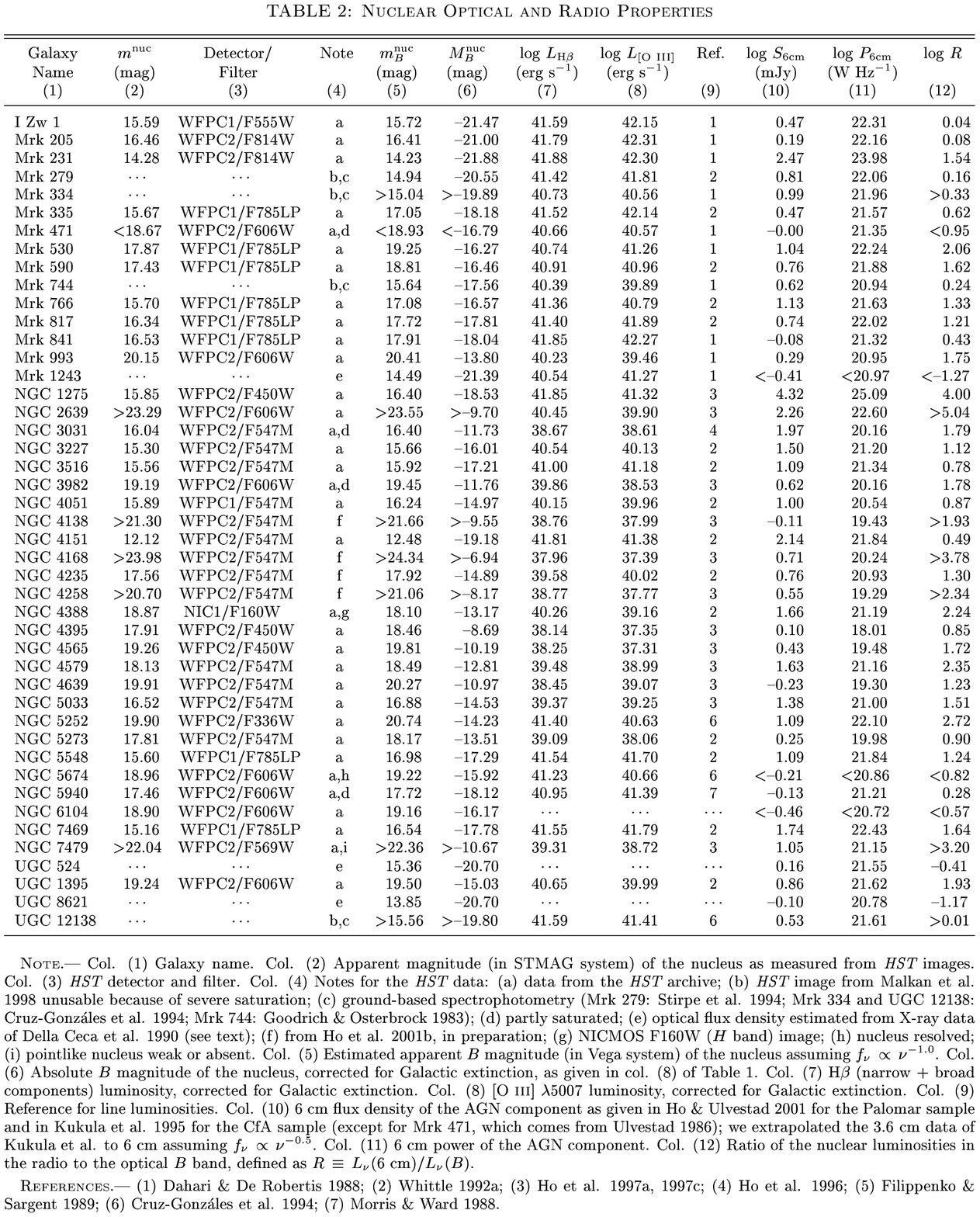,width=18.5cm,angle=0}}
\end{figure*}
 

\section{The Data}

\subsection{Palomar and CfA Seyfert 1 Nuclei}

The radio data come primarily from two sources.  Kukula et al. (1995) imaged 
the CfA sample at 3.6~cm (8.4~GHz) using the Very Large Array (VLA) in its A 
and C configurations; the resulting angular resolutions are $\sim$0\farcs25
and $\sim$2\farcs5, respectively.  They did not observe Mrk 471 because 
it was inadvertently omitted from the original list of Huchra \& Burg (1992). 
We used the data of Ulvestad (1986) for this object.  The radio properties of 
the full Palomar sample were recently investigated by Ho \& Ulvestad (2001).
They obtained scaled-array 6~cm (4.9~GHz) and 20~cm (1.4~GHz) continuum images 
with the VLA at a resolution of $\sim$1\asec.  Eighty-five percent of the 
objects were detected at 6~cm above a 3 $\sigma$ threshold of $\sim$0.12~mJy 
beam$^{-1}$.  The signal-to-noise ratio of the Kukula et al. maps is slightly 
lower than that of Ho \& Ulvestad, but the two surveys are otherwise quite 
compatible in terms of sensitivity and resolution.  Table~2 gives 6~cm 
monochromatic radio powers for the combined sample, derived from the observed 
flux densities assuming isotropic emission; the listed values pertain to the 
integrated emission from all components associated with the AGN.  The 3.6~cm 
data of Kukula et al. (1995) were extrapolated\footnote{Although 6~cm data 
have been published for many of the CfA Seyferts, we prefer to utilize the 
3.6~cm database of Kukula et al. (1995) for the sake of homogeneity.} to the 
fiducial wavelength of 6~cm assuming $f_{\nu}\,\propto\,\nu^{\alpha_{\rm r}}$, 
with $\alpha_{\rm r}\,=\,-0.5$, the median spectral index between 
6 and 20~cm found by Ho \& Ulvestad (2001).  We adopt the results of Ho \& 
Ulvestad (2001) for the few objects that overlap between the two samples.  
Altogether, 93\% (42/45) of the objects have radio detections, which range 
from $\sim$0.5~mJy to 21~Jy, with a median value of $\sim$6~mJy.

High-resolution ($\sim$0\farcs1) optical images for the majority (37/45) of the 
objects are available from the {\it Hubble Space Telescope}\ (\hst) studies of 
Nelson et al. (1996), Malkan, Gorjian, \& Tam (1998), and Ho et al. (2001b, 
in preparation), or from the \hst\ data archives.  Twenty-seven of the objects 
were observed with the Wide Field Planetary Camera 2 (WFPC2), and 10 with the
Wide Field Planetary Camera 1 (WFPC1), using a variety of medium-band and 
broad-band filters (see Table 2).  It is difficult to obtain accurate 
photometry of faint nuclei at optical wavelengths because the inner brightness 
distributions of bulges are both bright and sharply peaked (Phillips et al. 
1996; Carollo et al. 1997; Ho et al. 2001b, in preparation).  Even at \hst\ 
resolution, it is often nontrivial to disentangle the nucleus from the 
light of the inner bulge in the presence of complex fine structures 
(dust lanes, nuclear spirals, star clusters) invariably present, or to 
properly take into account the effects of the telescope point-spread function. 
All the \hst\ images, including those of Nelson et al. and Malkan et al. 
(neither of whom published photometry of the nuclei), were analyzed using 
GALFIT, a program for galaxy image decomposition developed by Peng et al. 
(2001, in preparation).  In brief, GALFIT fits to the observed image a 
two-dimensional model for the light distribution consisting of analytic 
functions for the bulge (``Nuker,'' S\'ersic, and exponential profiles; details
given in Peng et al. 2001, in preparation) plus an additional point source for 
the nucleus, convolved with an appropriate synthetic point-spread function. 
Applications of this method to \hst\ images of nearby galaxies are given in 
Ravindranath et al. (2001) and Ho et al. (2001b, in preparation).  

For the purposes of this study, we are interested only in obtaining 
luminosities for the nuclei, and we omit further discussion of the photometric
properties of the galaxy bulges.  The observed magnitudes are in the STMAG 
system, defined such that $m_{\rm STMAG}$ = $-$2.5 log $f_{\lambda}$ $-$ 21.1, 
where $f_{\lambda}$, in units of \flamb, is constant with wavelength (Holtzman 
et al. 1995).  To facilitate comparison within our sample, and also with 
quasar samples, we convert the magnitudes to the $B$ band in the Vega system, 
assuming $f_{\nu}\,\propto\,\nu^{\alpha_{\rm o}}$, with 
$\alpha_{\rm o}\,=\,-1.0$, a typical spectrum for the optical featureless 
continuum of Seyfert~1 nuclei (e.g., Ward et al. 1987).  The conversion 
factors were calculated using SYNPHOT within STSDAS in IRAF\footnote{IRAF is 
distributed by the National Optical Astronomy Observatories, which are 
operated by the Association of Universities for Research in Astronomy, Inc., 
under cooperative agreement with the National Science Foundation.}.

Nuclear magnitudes for the remaining eight objects were obtained from 
more heterogeneous sources, in the following manner.  The only \hst\ image 
available for NGC 4388 was taken with NICMOS in the F160W filter (1.6~\micron, 
$H$ band); this image was analyzed in the same manner as the optical 
images, as described by Ravindranath et al. (2001).  Ground-based optical 
spectrophotometry exists for four galaxies: Mrk 279, Mrk 334, Mrk 744, and 
UGC 12138.  Although not ideal (the data were typically acquired with 
2\asec--4\asec\ apertures), these observations do give a rough characteristic 
measurement of the central luminosity of the objects, which is certainly 
preferable to no correction at all for host-galaxy emission.  In any case, 
the contamination from the host, as judged by the depth of the stellar 
absorption lines, appears to be quite small for Mrk 279 and Mrk 744.  The 
optical flux densities were converted to $B$-band magnitudes, again using an 
$\alpha_{\rm o}\,=\,-1.0$ spectrum and SYNPHOT.  Finally, for three objects 
(Mrk 1243, UGC 524, and UGC 8621) we could only locate soft X-ray measurements 
from the catalog of Della Ceca et al. (1990).  We converted the observed 
count rates to a flux density at 2~keV assuming a power-law spectrum with an 
energy index of $\alpha_{\rm x}\,=\,-0.7$, as normally seen in type~1 Seyferts 
(e.g., Nandra et al.  1997).  Next, we estimated the near-ultraviolet flux 
density using a spectral index of $\alpha_{\rm ox}\,=\,-1.2$ between 2500~\AA\ 
and 2~keV, again a value typical of Seyfert~1s (Mushotzky \& Wandel 1989).  
And lastly, the $B$-band flux density was obtained by extrapolation from an
$\alpha_{\rm o}\,=\,-1.0$ spectrum.

A portion of our analysis will make use of optical emission-line luminosities.  
We utilize the \oiii\ \lamb 5007 and H\bet\ (narrow plus broad components) 
lines, the former to represent emission arising from the narrow-line region 
(NLR) and the latter from the NLR and broad-line region (BLR).  Both are 
presumed to be predominantly photoionized and hence trace reprocessed continuum 
emission.  In order to minimize systematic errors in the line luminosities due 
to aperture effects, we have restricted selection of the data to a few 
well documented, homogeneous sources.  Whenever possible we have given 
preference to the catalog of Whittle (1992a), who has carefully assembled 
reliable line measurements, followed by the Palomar catalog of Ho et al. 
(1997a).  Markarian Seyferts not included in Whittle's work were usually found 
in the compilation of Dahari \& De Robertis (1988).  

The photometric quantities of the sample are collected in Tables 1 and 2.  
The absolute magnitudes have been corrected for Galactic extinction but not 
for internal extinction.  In the case of the integrated magnitudes, we have 
refrained from correcting them to face-on inclination because in some cases 
the Hubble types and inclinations of the galaxies are poorly known.  And 
although measurements of the narrow-line Balmer decrement are available for 
most of the nuclei, we have chosen not to apply reddening corrections to the 
nuclear optical continuum and line luminosities.  The continuum light of the 
nucleus in general may experience a different amount of extinction compared to 
the emission-line gas.  In any case, the correction is modest and will not 
affect our main conclusions.  Using the data from Ho et al. (1997a), we find 
that the Palomar objects have a median Balmer decrement of H\al/H\bet\ = 3.7.  
For an intrinsic ratio of H\al/H\bet\ = 3.1 (Halpern \& Steiner 1983; Gaskell 
\& Ferland 1984) and the Galactic extinction law of Cardelli, Clayton, \& 
Mathis (1989), $A_B\,\approx$ 0.7 mag, which is comparable to other sources of 
systematic uncertainties (\S\ 3.3).

\subsection{Palomar-Green Quasars}

Our analysis will make frequent comparisons of the Seyfert sample with 
optically selected ``type~1'' AGNs of higher luminosity.   For this 
purpose we have chosen the well studied Palomar-Green (PG) sample of quasars 
(Schmidt \& Green 1983; Green, Schmidt, \& Liebert 1986).  We limit the 
PG sample to the 87 objects with $z\,<\,0.5$, since this subset has 
received the most detailed spectroscopic follow-up work.  The radio and 
optical continuum data come from Kellermann et al.  (1989).  
Boroson \& Green (1992)\footnote{The subsample of PG quasars in the study of 
Boroson \& Green (1992) was also independently investigated spectroscopically 
by Miller, Rawlings, \& Saunders (1993).  The study of Miller et al.  employed 
a wider slit than that of Boroson \& Green, so in principle the former 
achieved higher photometric accuracy than the latter, at the expense of 
somewhat degraded spectral resolution.  We have chosen to adopt Boroson \& 
Green's measurements because these authors accounted for the effects of 
contamination by Fe~II emission, which we deem to be important.  Miller et al. 
indicate that on average the \oiii\ fluxes of Boroson \& Green may
have been underestimated by 0.29 dex.} supply equivalent-width measurements 
for H\bet\ and \oiii; we obtained the corresponding line luminosities by 
combining the equivalent widths with the optical monochromatic continuum 
luminosities of Kellermann et al., properly extrapolated with 
$\alpha_{\rm o}\,=\,-1.0$. All distance-dependent quantities were transformed 
to our adopted cosmological parameters of $H_0$ = 75 \kms\ Mpc$^{-1}$ and 
$q_0$ = 0.5.

\subsection{Error Estimates}

We do not list errors explicitly for the photometric quantities because in 
most cases they are dominated by systematic uncertainties which are hard to 
quantify for each object individually.  Here we briefly assess their likely 
magnitude and their impact on our analysis.  Representative error bars appear 
in the figures later shown in the paper.  We assume, for simplicity, that 
distance errors can be neglected, although this is not well justified for the 
nearest galaxies whose distances can be rather uncertain.  

1. The formal errors of the radio flux densities are generally quite small, 
dominated by the $\sim$5\% uncertainty in the absolute flux scale (see, e.g., 
Ho \& Ulvestad 2001).  There is usually little ambiguity in assigning the 
proper component for the AGN.  For the most part, the radio images of 
Kukula et al. (1995) and Ho \& Ulvestad (2001) reveal relatively simple 
structures, usually consisting of a single, centrally dominant core, 
which is occasionally accompanied by linear features plausibly associated 
with plasma outflows.  The cores are compact and generally unresolved or 
barely resolved at resolutions of \lax 1\asec.  We explicitly assume that 
all of the emission from the compact core and linear features can be 
attributed to the AGN, with zero contribution from stellar sources.  In 
spiral galaxies, the integrated centimeter-wave radiation from \hii\ regions 
and supernova remnants can be significant, but on arcsecond scales the radio
emission of objects classified as \hii\ or starburst nuclei tends to be 
diffuse and usually extended along the optical disk (e.g., Condon et al. 1982; 
Filho, Barthel, \& Ho 2000).  One cannot rule out that the radio emission 
might arise from a compact starburst (e.g., Condon et al. 1991), but we regard 
this possibility to be unlikely for our objects since their optical spectra,
measured on comparable scales, by definition are classified as broad-lined 
AGNs, not starbursts.  Variability poses a potentially much more serious 
problem; however, little is known about the radio variability properties of 
Seyferts.  The handful of objects which have been monitored display intensity 
fluctuations at centimeter wavelengths ranging from $\sim$30\% (NGC 5548; 
Wrobel 2000) to $\sim$100\% (M81; Ho et al. 1999) over timescales of months.  
We choose 0.2~dex as a fiducial uncertainty for the radio powers.  Since the 
nuclear component of the 6~cm emission can be a significant fraction of the 
total (Fig.~1), this error estimate may be appropriate for both the nuclear 
and integrated powers.

2. Several sources of errors can potentially affect the nuclear continuum 
magnitudes.  (1) Even with \hst\ resolution, it is in practice very difficult 
to reliably measure nuclei fainter than $m_V\,\approx$ 19--20 mag superposed 
on the intrinsically cuspy cores of bright galaxies, because the inner light 
profile of the bulge blends smoothly with the weak point source.   The 
majority of the nuclei are brighter than this, however, and the simulations of 
Peng et al. (2001, in preparation) suggest that GALFIT can extract nuclear 
magnitudes to an accuracy of $\pm$0.2--0.3 mag for galaxies similar to the 
ones considered here.  (2) We use a single $f_{\nu}\,\propto\,\nu^{-1}$ 
spectrum to extrapolate all the continuum measurements to the $B$ band.  If we 
allow the optical power-law slope to vary by as much as 
$\Delta \alpha_{\rm o}\,=\,\pm 1$, sufficient to 


\vskip 0.3cm

\psfig{file=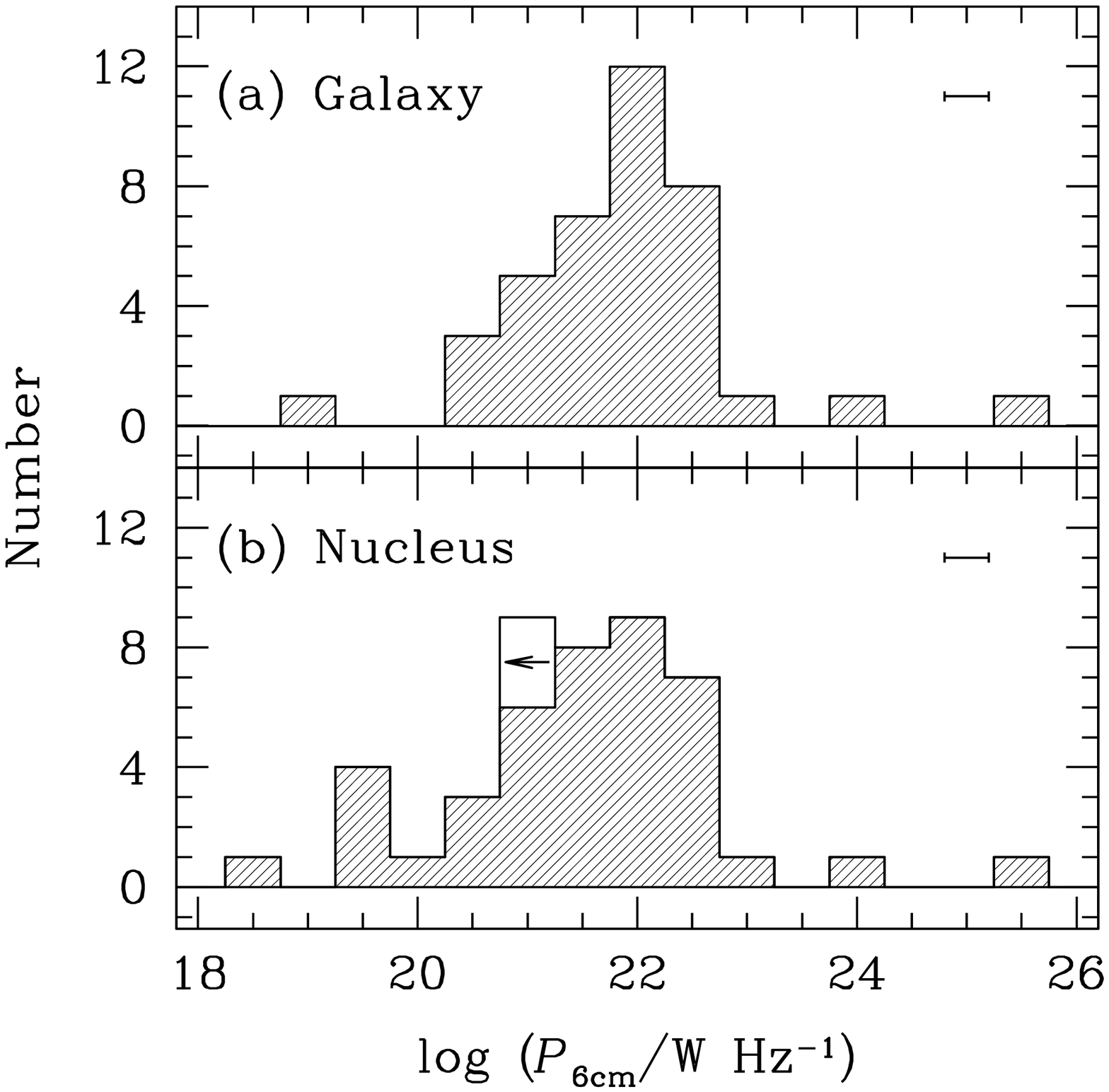,width=8.5cm,angle=0}
\figcaption[fig1.ps]{
Distribution of monochromatic 6~cm radio power for  ({\it a}) the integrated
emission of the galaxy and  ({\it b}) the AGN component alone.  The open
histograms denote upper limits. Typical error bars are shown (see \S\ 3.3).
\label{fig1}}
\vskip 0.3cm


\noindent
accommodate the most extreme 
spectral diversity observed (Elvis et al. 1994; Ho 1999b), the uncertainty in 
$m_B$ can be as much as $\sim$0.7 mag if the extrapolation is done from the 
$I$ band. More typically the observations were conducted in $V$, and if 
$\Delta \alpha_{\rm o}\,=\,\pm 0.5$, the uncertainty in $m_B$ is only 
$\sim$0.1 mag.  (3) As in the radio band, variability introduces an 
uncertainty which is hard to quantify, but monitoring studies of Seyfert 
nuclei find that long-term flux variations are generally \lax 0.5 mag (e.g., 
Hamuy \& Maza 1987; Winkler et al. 1992).  (4) For lack of a secure handle on 
the effects of dust, we do not account for dust extinction from the host 
galaxy or internal to the nuclear region.  Notwithstanding the ``type~1'' 
classification of our objects, the featureless continuum may be subject to 
nonnegligible extinction.   From their study of a hard X-ray selected sample of
Seyfert~1s, Ward et al. (1987) argue that visual extinctions of $\sim$1--3 
mag are not uncommon.  On the other hand, the conclusions of Ward et al. 
cannot be readily generalized to our optically selected sample, which may be 
biased toward less obscured sources.  Winkler (1997) obtains $A_V$ \lax 1 mag 
for the majority of a heterogeneous sample of 92 Seyfert~1s.  (5) Finally, as 
with the radio emission, we assume that the point source entirely originates 
from the AGN, with minimal contribution from starlight. Although this is most 
likely not the case for every object, especially for the few where only 
ground-based spectrophotometry was available, it is probably not an 
unreasonable assumption given the high angular resolution of the \hst\ 
images.   Taking the above factors into consideration, we conservatively 
assign an uncertainty of 1~mag to our $B$-band nuclear magnitudes.

3. The uncertainties of the emission-line luminosities mainly arise from 
errors in flux calibration, aperture effects, variability, and extinction.
The NLR extends over scales of at least tens to hundreds of parsecs, and 
therefore is partly resolved from the ground for nearby systems.  
Absolute spectrophotometry of extended objects is 


\vskip 0.3cm

\psfig{file=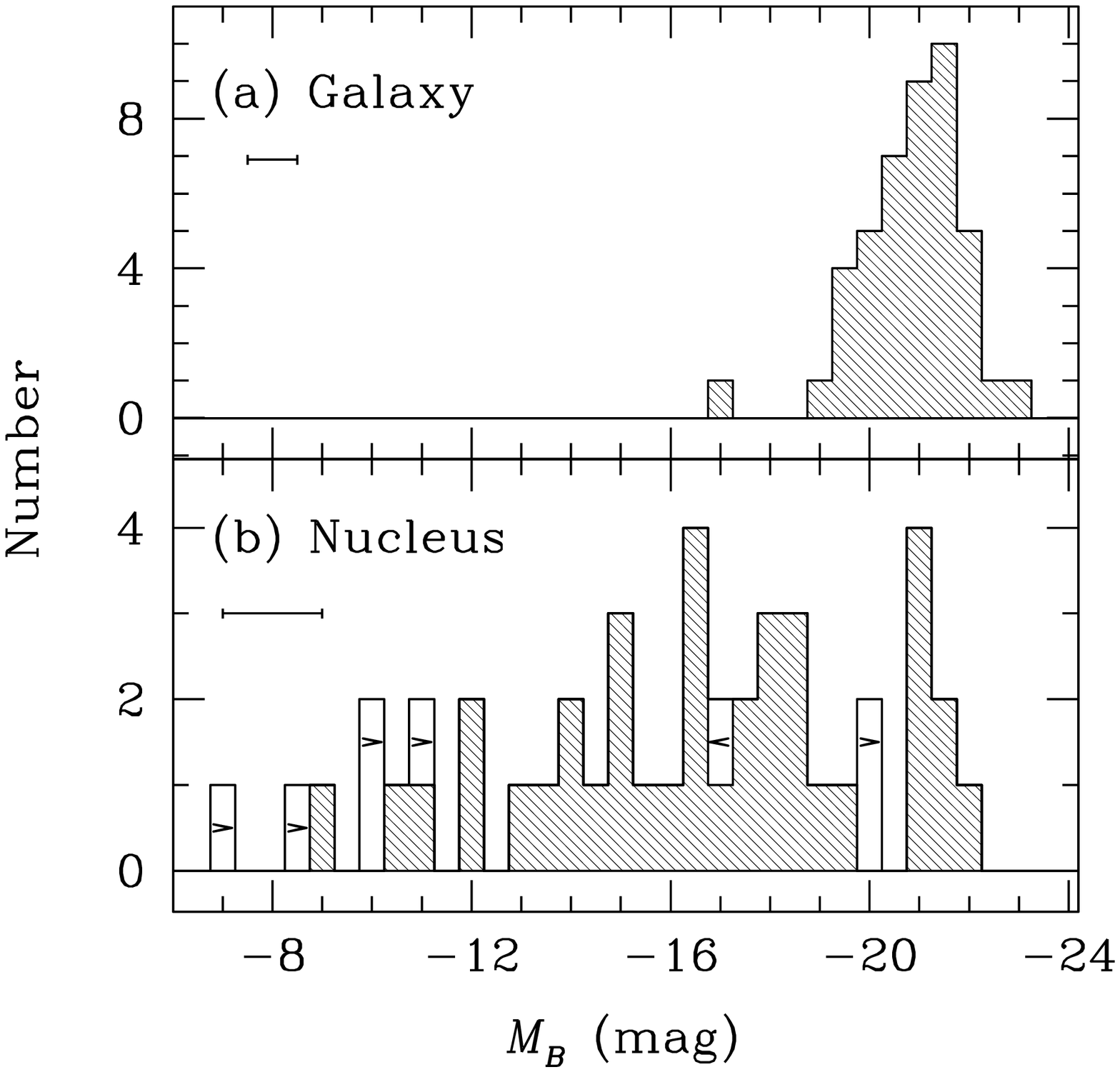,width=8.5cm,angle=0}
\figcaption[fig2.ps]{
Distribution of absolute $B$ magnitudes for ({\it a}) the integrated light of
the galaxy and ({\it b}) the nucleus alone.  The open histograms denote
limits.  Typical error bars are shown.
\label{fig2}}
\vskip 0.3cm


\noindent
notoriously challenging
without special effort, and aperture effects can be significant (see, 
e.g., discussion in Whittle 1992a).  Flux errors of a factor of $\sim$2 are 
not rare.  It is relatively straightforward to estimate the extinction along 
the line of sight to the NLR; for the Palomar sample, the median $A_B\,\approx$ 
0.6 mag (\S\ 3.1).  The situation for the BLR is more complicated (MacAlpine 
1985), and no generally accepted method exists to estimate the extinction
from optical spectra alone.  Lastly, the broad-line emission in Seyferts 
varies in response to changes in the continuum level.  We adopt a 
characteristic uncertainty of $\pm$0.3~dex for the \oiii\ and H\bet\ 
luminosities.

\section{Results}

\subsection{Radio and Optical Luminosities}

Figure~1 shows the distribution of monochromatic 6~cm radio powers.  The top 
panel plots the integrated emission from the whole galaxy (host plus nucleus), 
which we approximate with low-resolution (beam \gax 1\amin) measurements; 
the bottom panel isolates the nuclear contribution properly assigned to the 
AGN component.  The two distributions are not dramatically different 
statistically.  The AGN component on average accounts for $\sim$75\% 
of the integrated emission.  However, the differences for any individual
object {\it can}\ be very significant because the radio nuclear fraction 
varies from 0.01 to 1.  The nuclear 6~cm radio powers range from 
$\sim$10$^{18}$ to 10$^{25}$ W~Hz$^{-1}$, with a median value of 
$P_{\rm 6cm}$ = 1.6\e{21} W~Hz$^{-1}$, after accounting for the small number 
of upper limits (Feigelson \& Nelson 1985).  This distribution is quite 
similar to that of Ho \& Ulvestad (2001), who recently studied the entire 
sample of Seyferts (types~1 and 2) from the Palomar survey.  The level of 
emission seen here is also comparable to that of radio cores detected in a 
large fraction of nearby elliptical and S0 galaxies (Sadler, Jenkins, \& 
Kotanyi 1989; Wrobel \& Heeschen 1991; Ho 1999a).

The contrast between the nuclear and integrated light is much more dramatic at 
optical wavelengths (Fig.~2).  The 


\vskip 0.3cm

\psfig{file=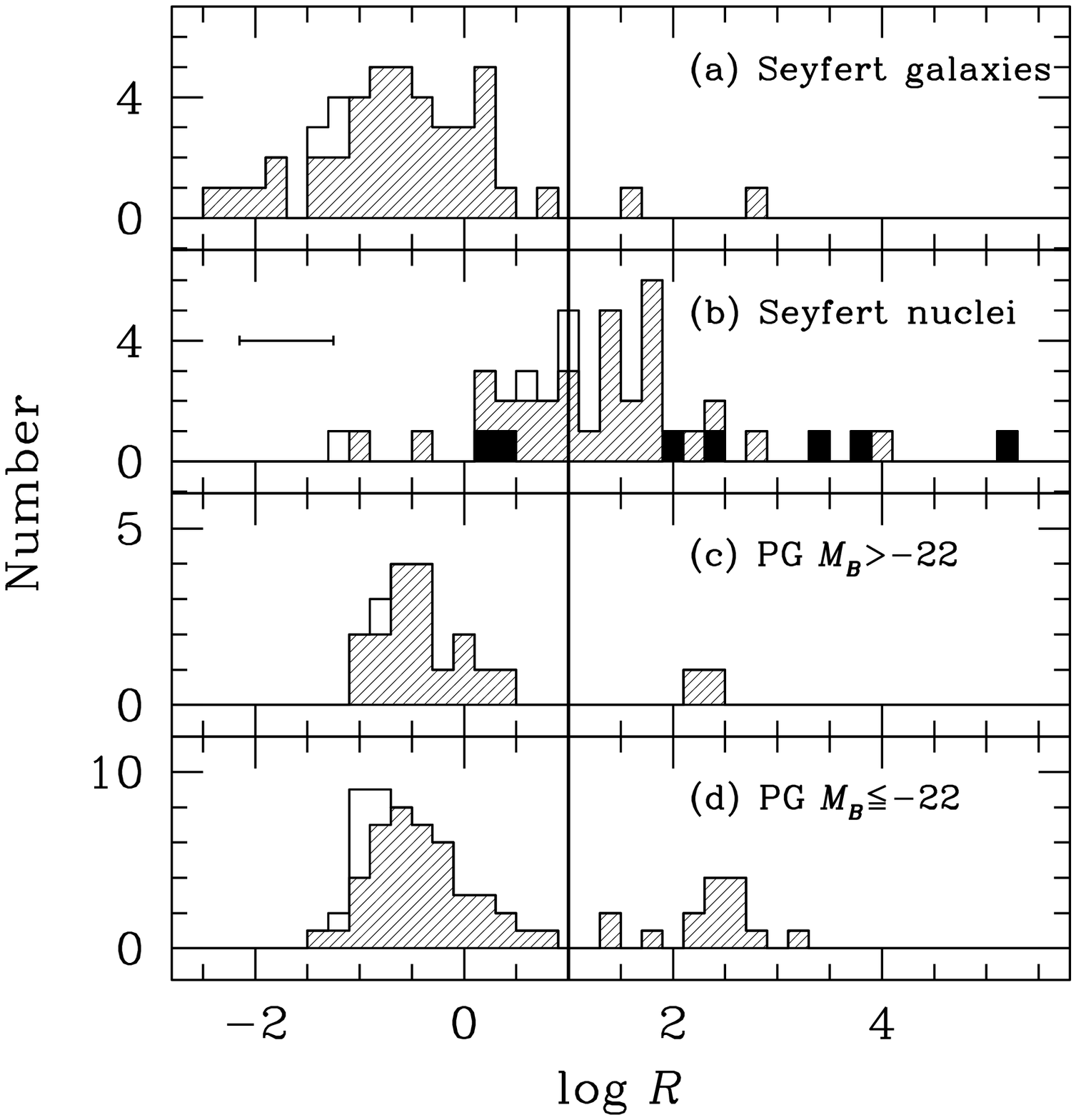,width=8.5cm,angle=0}
\figcaption[fig3.ps]{
Distribution of the radio-to-optical luminosity ratio $R$ for ({\it a}) the
entire Seyfert galaxy, ({\it b}) the Seyfert nucleus alone, ({\it c}) PG
sources with $M_B\,>\,-22$ mag, and ({\it d}) PG sources with
$M_B\,\leq\,-22$ mag.  Upper and lower limits are denoted by open and solid
histograms, respectively.  The formal division between radio-loud and
radio-quiet objects is given by $R$ = 10 ({\it solid vertical
line}).  A typical error bar for the Seyfert nuclei is shown in panel
({\it b}).
\label{fig3}}
\vskip 0.3cm


\noindent
median integrated absolute magnitude is 
$M_{B_T}\,=\,-20.8$, comparable to the valued obtained for all the Seyferts 
in the Palomar survey (Ho et al. 1997b).  The distribution of $M_{B_T}$ is 
rather narrow; the interquartile range is only 1.3 mag.  This reflects the 
fact that Seyfert nuclei inhabit a preferred, relatively restricted range of 
Hubble types.  In stark contrast, the nuclear magnitudes, properly decomposed 
from the bulge, span from $M_B^{\rm nuc}\,\approx\,-9$ to $-22$ mag, with no 
discernible peak in the distribution.  In the most extreme cases, the nucleus 
accounts for merely 0.01\% of the integrated light. The weakest Seyfert~1 
nuclei, which may be regarded as ``mini-quasars,'' are a factor of $\sim 10^8$ 
(20 magnitudes) less powerful than the most luminous quasars known --- a truly 
astonishing range for a single astrophysical phenomenon.  Treating the single
lower limit (Mrk 471, slight saturated) as a detection, the median value of 
$M_B^{\rm nuc}\,=\,-16.8$ mag, with an interquartile range of 4.7 mag.  
As expected, the CfA Seyferts are typically more luminous than the 
Palomar sources: median $M_B^{\rm nuc}\,=\,-17.4$ mag vs. $-14.6$ mag.  The 
two most luminous CfA Seyferts, I~Zw~1 and Mrk~231, have absolute magnitudes 
which approach the conventional, albeit arbitrary, border between Seyferts 
and quasars ($M_B\,=\,-22$ mag, adjusted to $H_0$ = 75 \kms\ Mpc$^{-1}$; 
Schmidt \& Green 1983).

\subsection{Radio-Loudness Parameter}

Our reevaluation of the radio-loudness parameter for Seyferts appears in 
Figure~3.  Not surprisingly, when $R$ is computed using the integrated 
optical and radio emission of the entire galaxy (Fig.~3{\it a}), the vast 
majority of the objects lie to the left of the boundary demarcating the 
radio-quiet/radio-loud regimes.  For a dividing line of 



\psfig{file=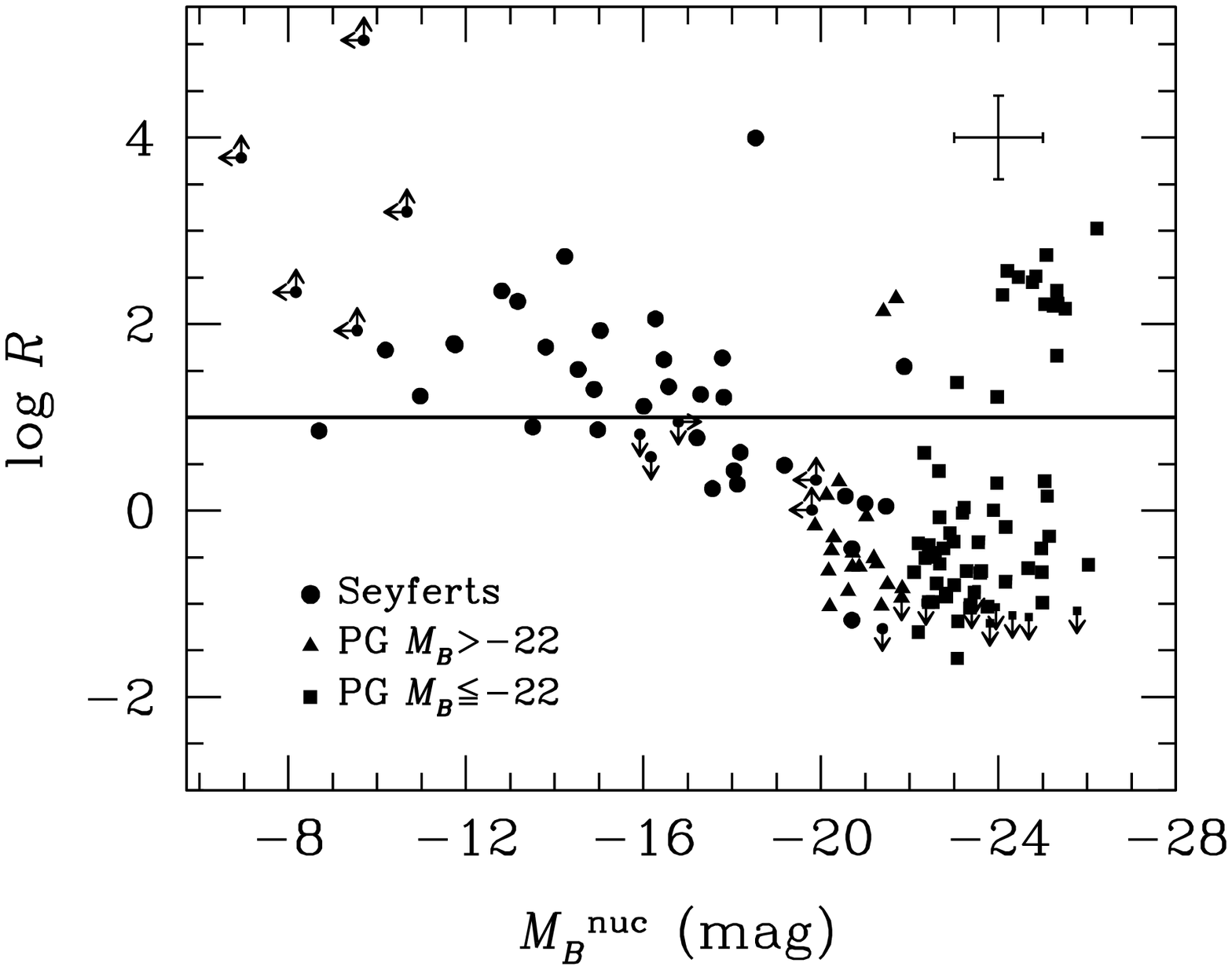,width=8.5cm,angle=0}
\figcaption[fig4.ps]{
Distribution of nuclear $R$ versus absolute $B$ magnitude.  {\it Circles}\
denote the Seyfert~1 nuclei from this study; {\it triangles}\ are PG sources
with $M_B\,>\,-22$ mag; and {\it squares}\ are PG sources with
$M_B\,\leq\,-22$ mag.  Limits are indicated with arrows.  The formal division
between radio-loud and radio-quiet objects is given by $R$ = 10 ({\it solid 
horizontal line}). A typical error bar for Seyfert nuclei is shown.
\label{fig4}}
\vskip 0.3cm


\noindent
$R$ = 10, the radio-loud fraction is 4\%; the median value is $R\,\approx$ 
0.2.  This is the result normally cited for Seyferts --- most or all Seyferts 
are radio quiet.  The values of $R$ calculated using high-resolution 
measurements of the nuclei are shown in Figure~3{\it b}.  Now the median value 
of $R\,\approx$ 17 (treating the seven lower limits as detections), and the 
fraction of the sample which technically qualifies as radio-loud sources 
changes to \gax 60\%.  The majority (75\%) of the objects lie between $R$ = 1 
and 100.  This is the main result of this paper.  

For comparison, we assembled equivalent data from the Palomar-Green (PG) 
quasar survey (Schmidt \& Green 1983), as compiled by Kellermann et al. 
(1989).  For consistency with our conventions, we recalculated the 
luminosities of the PG sources using $\alpha_{\rm r}\,=\,-0.5$,
 $\alpha_{\rm o}\,=\,-1.0$, $H_0$ = 75 \kms\ Mpc$^{-1}$, and $q_0$ = 0.5.  With 
these parameters, 23\% (20/87) of the PG objects have $M_B\,>\,-22$ mag and 
so can be deemed to be Seyfert~1 sources according to the definition of 
Schmidt \& Green.  As noted by Kellermann et al. (1989), the distribution of 
$R$ shows no marked difference between the PG Seyferts (Fig.~3{\it c}) and the 
{\it bona fide}\ quasars (Fig.~3{\it d}).  The PG sample, as do other 
optically selected quasar samples (e.g., Visnovsky et al. 1992; Hooper et al. 
1995; Goldschmidt et al. 1999) exhibits a deficit of objects in the region 
1\lax $R$ \lax 100, which leads to an apparent bimodal distribution of $R$.  
The local Seyfert nuclei, interestingly, occupy precisely the ``gap'' created 
by the PG sources.

Nearby type~1 AGNs, with characteristically lower optical luminosities,  
populate a unique region of parameter space compared to more distant, more 
luminous sources:  no object fainter than $M_B\,\approx\,-20$ mag has $R\,<$ 1 
(Fig.~4).  The absence of points on the lower left portion of the diagram 
cannot be the result of selection effects because of the high detection rate 
in the radio (93\%).  Instead, it reflects the fact the strength of the radio 
and optical emission are broadly correlated over a very wide range of 
luminosities, as shown below. 

\subsection{Relation between Radio and Optical Luminosities}

While the distribution of $R$ values for Seyfert nuclei is neither bimodal, as 
is the case for the PG objects, nor sharply peaked, for the majority of the 
objects it does span a relatively restrictive range (factor $\sim$100) over 
approximately four orders of magnitude in optical luminosity.  This suggests 
that a correlation should exist between the radio and optical continuum 
luminosities.  As shown in Figure~5{\it a}, a loose relation indeed is present 
between these two quantities.  Interestingly, radio-loud and radio-quiet 
Seyferts lie on two slightly offset sequences, the extrapolations of which 
merge with similar sequences for the PG quasars.  The radio-loud branch is 
steeper than the radio-quiet branch.  Repeating the analysis by substituting 
the $B$-band continuum magnitudes with the \oiii\ (Fig.~5{\it b}) and H\bet\ 
(Fig.~5{\it c}) luminosities yields qualitatively very similar results.

Luminosity-luminosity correlations of the type shown in Figure~5 can be 
potentially misleading, since each of the variables itself correlates 
strongly with distance.  A common practice is to reconsider the correlations 
in flux-flux space.  As discussed by Feigelson \& Berg (1983), however, such 
a procedure can lead to ambiguous results.  Unless the two luminosities are 
linearly correlated, and unless upper limits are few or absent, the flux-flux 
plane may exhibit little or no correlation even if an intrinsically strong 
correlation is present in the luminosity-luminosity plane.  The solution to 
safeguard against spurious distance effects is to perform a partial 
correlation analysis, taking distance as the third variable.  In the case of 
censored data sets, Akritas \& Siebert (1996) proposed the partial Kendall's 
$\tau$ correlation test.  Applying this test to our data, we verified 
that the statistical significance of the radio-optical correlations is high 
($>$99.9\%; see Table~3).  The correlations hold for Seyferts and quasars 
combined, for Seyferts alone, or for just radio-quiet quasars alone.   The 
radio-loud quasars by themselves are not significantly correlated, probably
because of the small number of objects.

To quantify the radio-optical relations, we calculated the linear regression 
coefficients using the method of Schmitt (1985), which treats censoring in 
both variables.  The results are reported in Table~3.   In Figure~5, we 
plot the regression lines separately for radio-loud and radio-quiet 
objects.  Representing $L_{\rm radio}\,\propto\,L^a_{\rm optical}$, 
$a\,\approx$ 1.1--1.3 for radio-loud sources, and $a\,\approx$ 0.50--0.70 
or radio-quiet sources.

\vskip 0.3cm

\subsection{Optical Continuum and Emission-Line Luminosity Correlation}

Among the lines of evidence supporting the physical continuity between quasars 
and ``classical'' Seyfert nuclei, two historically important ones are (1) that 
the optical luminosities of the two classes overlap smoothly (Weedman 1976) 
and (2) that in both classes the optical nonstellar emission correlates 
strongly with the Balmer emission-line luminosity (Yee 1980; Shuder 1981).  
The latter observation, in particular, was instrumental in establishing 
photoionization as the principal mechanism for powering the optical line 
emission.   Figure~6 revisits the relation between optical continuum and 
Balmer-line emission for type~1 AGNs.  The PG sources follow a well defined 


\vskip 0.3cm


\begin{figure*}[t]
\centerline{\psfig{file=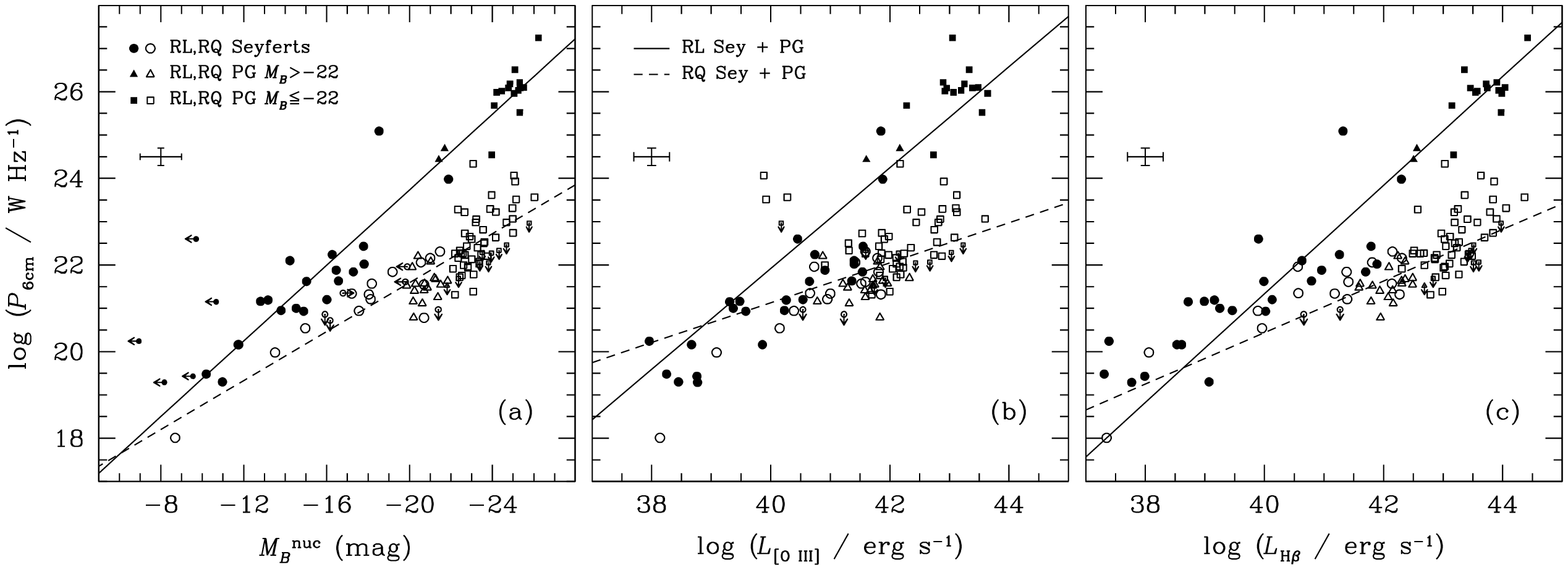,width=19.5cm,angle=270}}
\figcaption[fig5.ps]{
Correlations between nuclear monochromatic 6~cm radio power and ({\it a})
absolute $B$ magnitude, ({\it b}) \oiii\ \lamb 5007 luminosity, and ({\it c})
H\bet\ (broad + narrow components) luminosity.  {\it Circles}\ denote
the Seyfert~1 nuclei from this study; {\it triangles}\ are PG sources with
$M_B\,>\,-22$ mag; and {\it squares}\ are PG sources with
$M_B\,\leq\,-22$ mag.  Radio-loud and radio-quiet sources are plotted with
filled and open symbols, respectively, and arrows indicate limits.  Significant
Table 3) are shown for radio-loud ({\it solid line}) and radio-quiet ({\it
dashed line}) objects.  Typical error bars for Seyfert nuclei are shown.
\label{fig5}}
\end{figure*}
\vskip 0.3cm


\noindent
relation between H\bet\ luminosity and absolute $B$-band magnitude, roughly 
over 41.5 $<\,\log~L_{{\rm H}\beta}\,<$ 44.5 and $-20$ mag $\,<\,M_B\,<\,-26$ 
mag.  The scatter is small, but nonnegligible, as discussed by Miller et al. 
(1992).  Albeit with considerably greater scatter, the local Seyferts extend 
the correlation to $\log~L_{{\rm H}\beta}\,\approx\,38$ and $M_B\,\approx\,
-8.0$ mag, approximately with the same slope and intercept defined by the 
quasars (see Table~3).  The relation for Seyferts ($L_{{\rm H}\beta}\,\propto\,
L_B^{0.65}$) appears marginally shallower than that for quasars 
($L_{{\rm H}\beta}\,\propto\, L_B^{0.93}$), but we are unsure of the
significance of the difference.  The detection of broad emission lines, 
on which the type~1 Seyfert classification is based, becomes increasingly 
challenging for very low-luminosity objects (see Ho et al. 1997c).  Thus, 
it is possible that the distribution of points toward the faint end of the 
$L_{{\rm H}\beta}-M_B$ relation is biased toward higher values of 
$L_{{\rm H}\beta}$, hence flattening the slope.  Note that stellar 
contamination has the opposite effect: if the true nonstellar continuum in the 
Seyfert sample is less than we assume, the slope of the 
$L_{{\rm H}\beta}-L_B$ relation would be even flatter.  Radio-loud and 
radio-quiet objects lie on the same relation.   The increased scatter for the 
low-luminosity sources may be largely due to the difficulty of the 
measurements (\S\ 3.3), but it could partly reflect {\it intrinsic}\ 
variations of physical conditions with luminosity.  Two obvious possibilities 
are differences in the covering factor of the line-emitting gas and the 
spectral energy distribution.  Recent work suggests that the broad-band 
spectra of low-luminosity AGNs deviate strongly from those of high-luminosity 
sources (Ho 1999b; Ho et al. 2000c).  

As with the radio-optical relations (\S\ 4.3), the $L_{{\rm H}\beta}-M_B$ 
correlation is statistically sound.  After accounting for the mutual 
dependence of $L_{{\rm H}\beta}$ and $M_B$ on distance, the probability for 
accepting the null hypothesis that the two variables are uncorrelated is 
7\e{-4} for Seyferts and $< 10^{-8}$ for PG quasars (Table~3).  The 
best-fitting linear regression line for the combined sample, calculated using 
Schmitt's (1985) method, is 

$$\log L_{{\rm H}\beta}\,=\,(-0.34\pm 0.012) M_B\,+\,(35.1\pm 0.25).$$

\vskip 0.6cm

\section{Discussion}

Active nuclei are commonplace in nearby galaxies, but they are nontrivial 
to study quantitatively.  This paper demonstrates the importance of angular 
resolution in observations of Seyfert nuclei at radio and optical wavelengths.
With its unprecedently broad luminosity coverage, our sample of objects pushes 
the Seyfert phenomenon to unfamiliar territories and leads us to reexamine 
a few longstanding issues.

\subsection{The Radio-Loudness Paradigm}

The physical processes underpinning the generation of radio sources remain 
central topics of discussion in the AGN community.  Of particular interest is 
the origin of the apparent radio-loud/radio-quiet dichotomy.  As discussed 
in the Introduction, Seyfert nuclei traditionally have been considered 
radio-quiet AGNs.  This work emphasizes the difficulties that arise when 
trying to apply to Seyfert galaxies the conventional radio-loudness 
criterion.  The standard measure of the relative luminosity in the radio and 
optical band is only meaningful when the photometric measurements pertain to 
the active nucleus.  In the case of Seyferts, the nucleus often is dwarfed by 
the integrated emission from the host galaxy.  

Our primary result, the distribution of nuclear $R$ values for Seyfert~1 
nuclei, challenges three pieces of popular wisdom: (1) that Seyfert 
galaxies are mainly radio-quiet objects; (2) that radio-loud AGNs are seldom 
found in disk galaxies; and (3) that the majority of AGNs are radio quiet.  We 
find that at least 60\% of our sample 


\vskip 0.3cm

\psfig{file=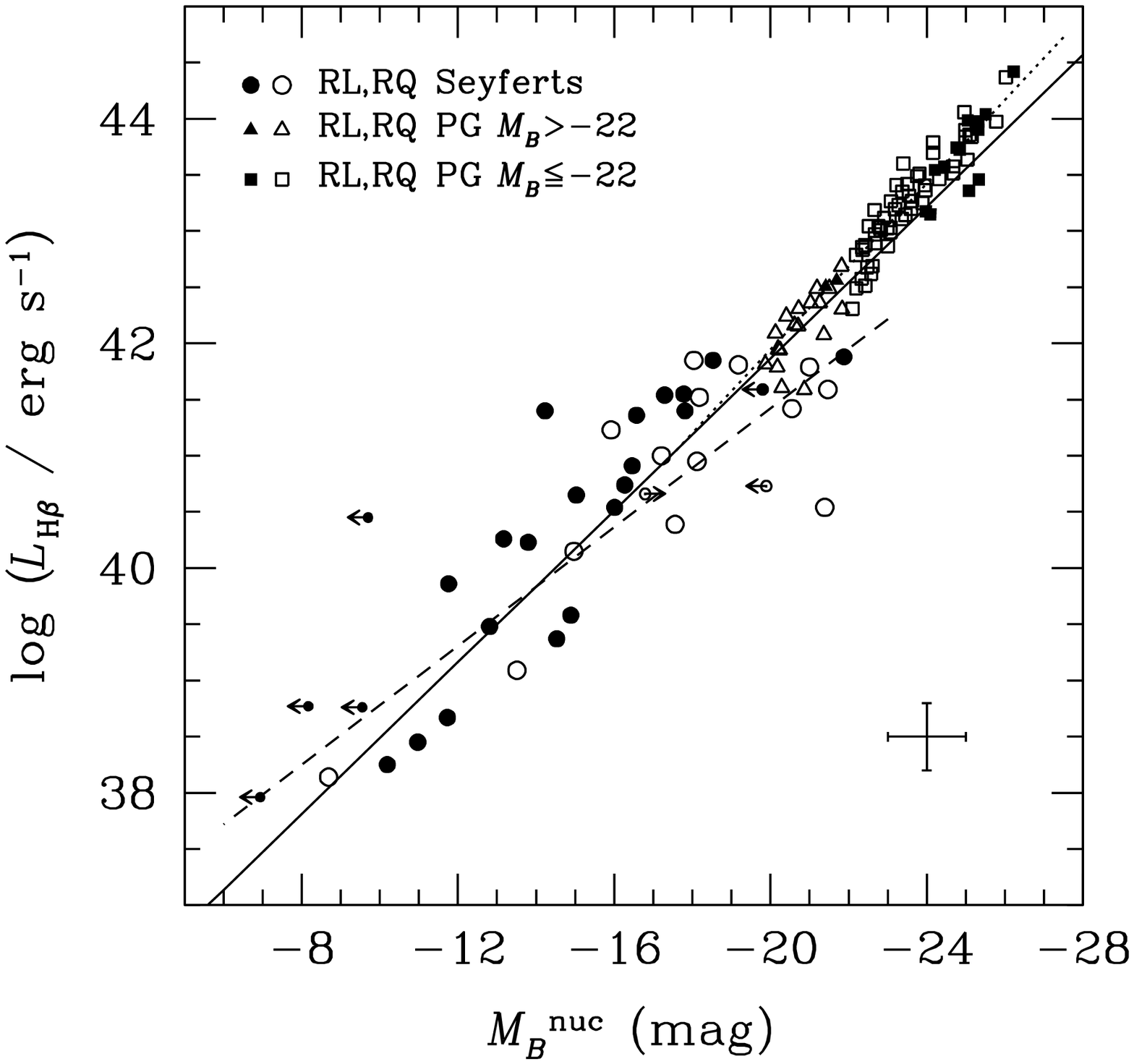,width=8.5cm,angle=0}
\figcaption[fig6.ps]{
Correlation between nuclear H\bet\ (broad + narrow components) luminosity
and absolute $B$ magnitude.  Symbols as in Figure 5.  The best-fitting
linear regressions are shown separately for all the objects ({\it solid
line}), for the Seyferts alone ({\it dashed line}), and for the PG objects
alone ({\it dotted line}).  A typical error bar for Seyfert nuclei is shown.
\label{fig6}}
\vskip 0.3cm


\noindent
of Seyferts technically qualify as 
radio-loud objects, and as Table~1 indicates, nearly all of them are spiral 
or S0 galaxies.  The existence of radio-loud nuclei in disk galaxies has been 
discussed in related contexts by Ho (1999b) and Ho et al. (2000c).  These 
remarks are not meant to undermine genuine differences that might exist 
between radio sources in Seyferts compared to those in more powerful AGNs.
Most Seyferts have $R\,\approx\,1-100$, values that straddle the two 
peaks in the $R$ distribution of optically selected quasars, and they are 
certainly much less extreme than the $R$ values seen in radio selected AGNs.  
Another feature which distinguishes Seyferts from typical radio-loud AGNs 
is the virtual absence of well collimated, large-scale jets.  Seyferts do
often possess linear radio structures reminiscent of jetlike outflows, but 
they occur exclusively on sub-galactic (\lax 1 kpc) scales (e.g., Ulvestad 
\& Wilson 1989; Kukula et al. 1995; Ho \& Ulvestad 2001).  The scant available 
measurements of jet speeds suggest sub-relativistic motions (Ulvestad et al. 
1998, 1999), although the first instance of superluminal jet motion in a 
Seyfert galaxy was recently reported by Brunthaler et al. (2000).

It is worth noting that although our definition of the radio-loudness 
parameter is widely used in the literature, it is by no means universally 
accepted.  Miller et al. (1990) argue that the radio power alone should be 
regarded as a more fundamental discriminant.  Their limiting criterion,
$P_{\rm 6cm}\,\approx\,10^{25}$ W~Hz$^{-1}$ sr$^{-1}$, would exclude all 
but one of the objects in our sample, the well known radio source NGC 1275 
(3C~84).  However, we hold the perspective that the radio emission has 
physical significance in relation to the other sources of energy emanating 
from the central engine, and we consider the $R$ parameter, which captures the 
{\it relative}\ radiative output in the radio band, to be meaningful.\footnote{
Rush et al. (1996) and Thean et al. (2000) have evaluated the radio-loudness 
of Seyfert galaxies according to the relative luminosities in the radio 
and far-infrared bands.  The far-infrared data, however, are based on 
low-resolution measurements obtained with {\it IRAS}, and thus suffer 
severe contamination from the host galaxy.}

\subsection{Radio Emission and Accretion Rate}
The connection between radio and optical emission has been much discussed 
in the context of quasars and powerful radio galaxies.  The most widely 
reported correlations are those between radio power and optical narrow-line
luminosity (e.g., Baum \& Heckman 1989; Rawlings et al. 1989; Rawlings \& 
Saunders 1991; Miller et al. 1993; McCarthy 1993; Tadhunter et al. 1998; 
Xu, Livio, \& Baum 1999), broad-line luminosity (Celotti, Padovani, \& 
Ghisellini 1997; Cao \& Jiang 1999, 2001), and optical continuum luminosity 
(Miller et al. 1990; Stocke et al. 1992; Lonsdale, Smith, \& Lonsdale 1995; 
Kukula et al. 1998; Serjeant et al. 1998).  

We find that radio-quiet PG quasars obey strong correlations between radio 
power and three separate measures of optical luminosity ($M_B$, 
$L_{\rm [O~III]}$, $L_{{\rm H}\beta}$), independent of distance effects.  This 
confirms qualitatively similar findings in previous investigations of the 
PG sample (Stocke et al. 1992; Miller et al. 1993; Lonsdale et al. 1995).   
Radio-loud PG quasars do not follow these correlations, most likely because of 
the small sample size and its limited dynamic range in luminosity. 

Seyfert nuclei form a natural extension of quasars in the radio-optical 
luminosity diagrams.  Moreover, with our definition of the radio-loudness 
parameter, each of two radio classes plausibly follows its own correlation. 
Of the three radio-optical correlations for Seyferts, the best known is that 
involving the \oiii\ luminosity (de~Bruyn \& Wilson 1978; Whittle 1985, 
1992b; Giuricin, Fadda, \& Mezzetti 1996).  Our results improve on the 
previous studies in several respects.  First, our $P_{\rm 6cm}-L_{\rm [O~III]}$ 
diagram has virtually no upper limits.  Second, its scatter is considerably 
reduced, most likely because of the high detection rate, the homogeneity of 
our sample, and the fact that we have restricted ourselves only to type~1 
sources.  Third, we (as do Giuricin et al.) explicitly account for distance 
effects.  And fourth, we expand on the dynamic range of the luminosities, 
not only toward weaker sources but also connecting with the quasar domain.

The radio power correlates with H\bet\ luminosity with equal, and possibly 
even better, significance compared to $L_{\rm [O~III]}$.  Baum \& Heckman 
(1989) and Ho (1999a) have noted that Seyferts do exhibit a loose 
correlation between radio and H\al\ luminosity, but that their locus 
lies offset from that formed by powerful radio galaxies.  By contrast, the
``radio-loud'' Seyferts in our sample are not inconsistent with a simple 
extrapolation of the radio-loud PG quasars (Fig.~5{\it c}).

Published studies of the relationship between the radio and optical continuum 
luminosities of Seyferts give mixed, and at times conflicting, accounts.  
The integrated optical luminosities do correlate with radio power (Isobe, 
Feigelson, \& Nelson 1986; Edelson 1987; Giuricin et al. 
1990), but evidently {\it not}\ when distance effects are considered (Giuricin 
et al. 1996).  Giuricin et al. (1996) collected ground-based nuclear 
magnitudes for a 


\begin{figure*}[t]
\centerline{\psfig{file=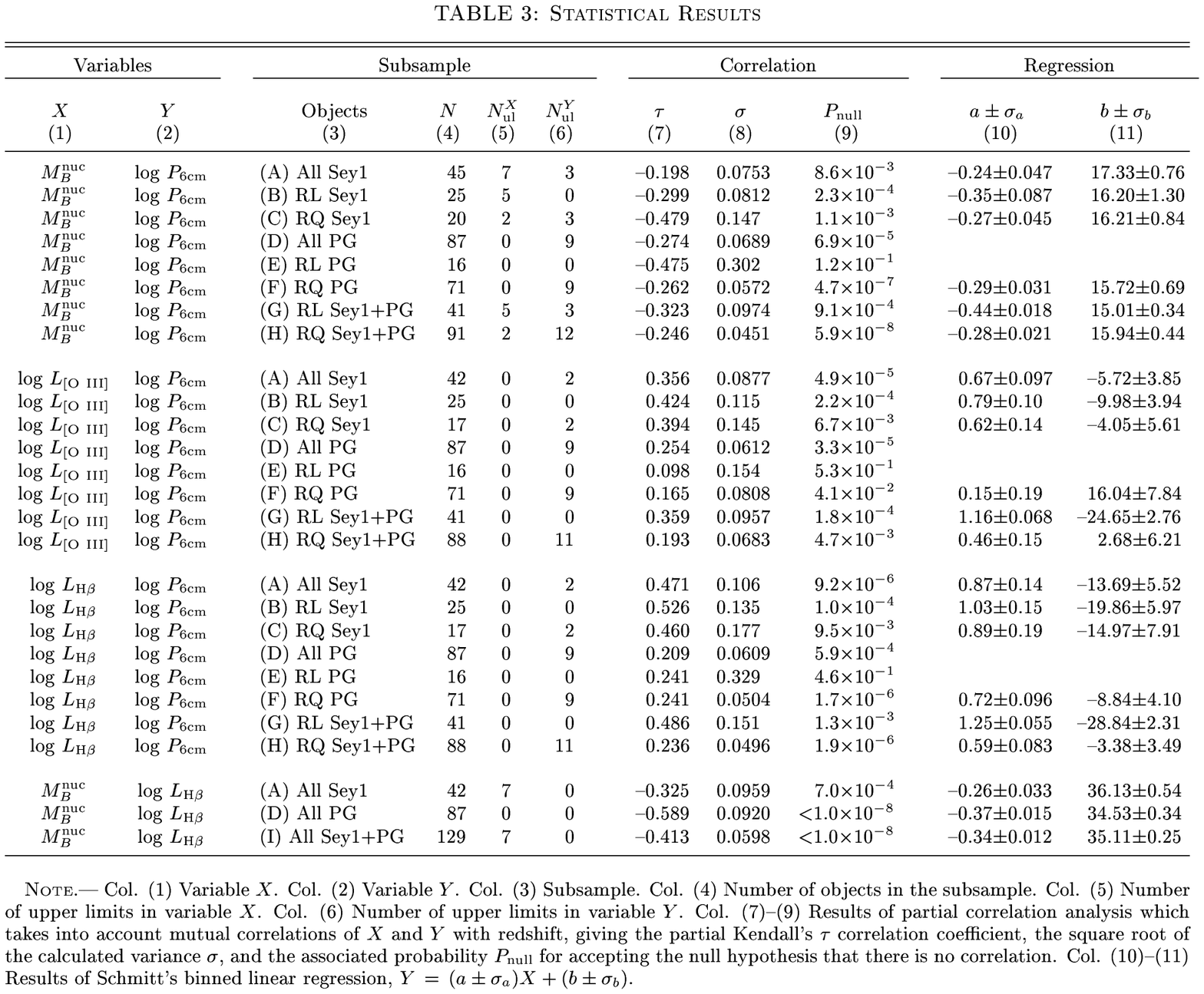,width=18.5cm,angle=0}}
\end{figure*}


\noindent
heterogeneous sample of Seyfert~1 nuclei, and these, too, 
were found to be intrinsically decoupled from the radio powers.  Our sample, on 
the contrary, shows a strong, intrinsic correlation between $P_{\rm 6cm}$ and 
$M_B$ (Fig.~5{\it a}).  In fact, we suggest that separate $P_{\rm 6cm}-M_B$  
relations can be defined for each of the two radio classes of Seyferts and 
quasars combined.  The discrepant results of Giuricin et al. (1996) perhaps 
stem from the large fraction of upper limits in their sample ($\sim$60\% for 
both radio and optical luminosities), the inhomogeneous sources of published 
measurements they used, or residual contamination of host-galaxy emission in 
their ground-based optical data.  That Seyferts and radio-quiet quasars share a 
single radio-optical continuum correlation was first proposed by Kukula et al. 
(1998), who compared the Seyfert~1s from the CfA sample with a subset of 27
low-redshift PG sources.  To mitigate host-galaxy contamination for the 
Seyferts, Kukula et al. utilized ground-based ``nuclear'' magnitudes 
from Huchra \& Burg (1992).  Two caveats, however, can be raised against 
the analysis of Kukula et al.  First, they did not consider the effect of 
distance on the correlation.  And second, as these authors recognized, 
their correction for host-galaxy light yields only crude AGN magnitudes, 
since Huchra \& Burg's photometry was based on apertures of diameters 
\gax 10\asec--15\asec.

Although the empirical case for the reality of the various radio-optical 
correlations seems secure, the physical basis behind them is less transparent.
In the case of Seyferts, the correlations between radio power and both the 
kinematics (Wilson \& Willis 1980; Whittle 1985) and luminosity (see above) 
of the NLR have long implicated a strong coupling between the energetics of 
the relativistic plasma and the narrow-line gas (de~Bruyn \& Wilson 1978).  
The correlations presented in Figure~5 inject a fresh perspective into the 
discussion.  The radio power does not correlate {\it just}\ with the NLR 
(\oiii) luminosity, but at least as well with the nonstellar continuum and the 
BLR luminosity (included in $L_{{\rm H}\beta}$).  In fact, all three 
radio-optical correlations exhibit approximately the same form and scatter.  
This suggests that there may be a single underlying cause.  

The simplest interpretation is that the accretion rate directly influences the 
fueling of relativistic radio jets, as invoked in certain models (e.g., Falcke 
\& Biermann 1995; Xu et al. 1999, and references therein).  Among the three 
measures of the optical luminosity we have considered, the nonstellar 
continuum may be regarded as the most ``fundamental.''  We assume, as is 
customary, that the featureless optical continuum represents the low-energy 
tail of the ionizing spectrum produced by the central accretion disk.  
Photoionization then powers the BLR and NLR emission, which we measure by 
H\bet\ and \oiii.  Indeed, the $L_{{\rm H}\beta}-M_B$ relation (\S\ 4.4) 
supports this scenario.  Thus, to the extent that $M_B$, $L_{\rm [O~III]}$, 
and $L_{{\rm H}\beta}$ each traces the accretion luminosity, which scales 
linearly with the mass accretion rate for an optically thick, geometrically 
thin disk, all three optical luminosities are equivalent.

\vskip 0.3cm

\subsection{Low-Luminosity AGNs}
Although the physical nature of Seyfert nuclei normally is not 
controversial, some of the objects in our sample have such extremely 
low luminosities that their status as ``AGNs'' may seem questionable.  
Indeed, the objects at the faint end of the luminosity distribution are 
no brighter than single supergiant stars.  Apart from the 
presence of a BLR, as implied by the detection of broad emission lines, this 
study, in conjunction with that of Ho \& Ulvestad (2001), furnishes several
other pieces of evidence which help to establish a physical continuity between 
low-luminosity Seyfert nuclei and more luminous AGNs. (1) Compact radio cores 
are detected nearly ubiquitously, often accompanied by linear features 
morphologically akin to radio jets.  (2) Judging from their radio-to-optical 
luminosity ratios, a substantial fraction of the sources are ``radio-loud,'' 
a trait normally associated with the AGN phenomenon.  (3) The radio power 
scales with three measures of the optical power (continuum, \oiii, and H\bet\ 
emission) and merges continuously with the quasar population.  (4) Lastly, 
a strong relation exists between optical continuum luminosity and H\bet\ 
luminosity, covering nearly seven orders of magnitude in each variable.
Terashima, Ho, \& Ptak (2000) and Ho et al. (2001a), using hard X-ray 
observations, recently arrived at similar conclusions for related classes of 
low-luminosity nuclei.  These findings support the hypothesis that AGNs with a 
remarkably broad range of power, ranging from quasars to classical Seyferts to 
ultra low-luminosity nuclei, are regulated by similar basic physical processes.  

\section{Summary}

We present radio and optical photometry for the nuclear regions of a well 
defined sample of 45 nearby Seyfert~1 galaxies to investigate the relationship 
between the emission in these two bands.  We emphasize the importance of high 
angular resolution to properly isolate the AGN component from the 
contaminating bright background of the host galaxy.  These data are 
supplemented with similar measurements for the subset of 87 Palomar-Green 
quasars with redshifts less than 0.5, allowing us to to assemble a sizable 
sample of type~1 AGNs which covers nearly seven orders of magnitude in radio 
and optical luminosity.  Our principal conclusions are the following:

\begin{itemize}

\item{
The nuclear radio-to-optical luminosity ratios of Seyfert~1 nuclei largely
range from $R\,\approx$ 1 to 100.  At least 60\% of the sources are
characterized by $R\,\geq\,10$, a territory traditionally reserved for
radio-loud AGNs.  This finding dispels the popular notion that most
Seyferts, and by implication their spiral hosts, are radio-quiet objects.
}

\item{ 
Seyfert~1 nuclei and optically selected quasars collectively display
strong, intrinsic correlations between radio and optical power, the latter
including nonstellar continuum, \oiii, and H\bet\ luminosity.  Separate
correlations can be defined for radio-loud and radio-quiet sources.
}

\item{ 
The physical origin underlying the radio-optical correlations may be a
close link between the disk accretion rate and the generation of relativistic
radio jets.
}

\item{
The relation between optical continuum and Balmer-line luminosity
extends from $M_B\,\approx\,-26$ mag to $M_B\,\approx\,-8$ mag.
}

\item{ 
The physical processes responsible for the production of radio and optical
emission in low-luminosity Seyfert nuclei are closely related to those
in classical Seyferts and quasars.
}
 
\end{itemize}

\acknowledgements
We thank J.~S. Ulvestad, A.~V. Filippenko, and W.~L.~W. Sargent for permission 
to use data in advance of publication, J.~J. Condon, the referee, for helpful 
comments, and D.~L. Meier for insights on jet physics.  The research of 
L.~C.~H. is supported by NASA grants HST-GO-06837.04-A, HST-AR-07527.03-A, and 
HST-AR-08361.02-A, awarded by the Space Telescope Science Institute, which is 
operated by AURA, Inc., under NASA contract NAS5-26555.  This work made use of 
the NASA/IPAC Extragalactic Database (NED) which is operated by the Jet 
Propulsion Laboratory, California Institute of Technology, under contract with 
NASA.


\begin{thebibliography}{}

\bibitem[]{} 
Adams, T.~F. 1977, \apjs, 33, 19

\bibitem[]{} 
Akritas, M.~G., \& Siebert, J. 1996, \mnras, 278, 919

\bibitem[]{} 
Baum, S.~A., \& Heckman, T.~M. 1989, \apj, 336, 702

\bibitem[]{} 
Becker, R.~H., White, R.~L., \& Edwards, A.~L. 1991, \apjs, 75, 1

\bibitem[]{} 
Blandford, R.~D. 1990, in Active Galactic Nuclei, SAAS-FEE Advanced Course 20,
Swiss Society for Astrophysics and Astronomy, ed. T.~J.-L. Courvoisier \&
M. Mayor (Berlin: Springer), 161

\bibitem[]{} 
Boroson, T.~A., \& Green, R.~F. 1992, \apjs, 80, 109

\bibitem[]{} 
Brunthaler, A., Falcke, H., Bower, G.~C., Aller, M.~F., Aller, H.~D.,
Ter\"{a}sranta, H., Lobanov, A.~P., Krichbaum, T.~P., \& Patnaik, A.~R. 
2000, \aa, 357, L45

\bibitem[]{} 
Cao, X., \& Jiang, D.~R. 1999, \mnras, 307, 802

\bibitem[]{} 
------. 2001, \mnras, 320, 347

\bibitem[]{} 
Cardelli, J.~A., Clayton, G.~C., \& Mathis, J.~S. 1989, \apj, 345, 245

\bibitem[]{} 
Carollo, C.~M., Stiavelli, M., de Zeeuw, P.~T., \& Mack, J. 1997, \aj,
114, 2366

\bibitem[]{} 
Celotti, A., Padovani, P., \& Ghisellini, G. 1997, \mnras, 286, 415

\bibitem[]{} 
Condon, J.~J. 1987, \apjs, 65, 485

\bibitem[]{} 
Condon, J.~J., Condon, M.~A., Gisler, G., \& Puschell, J.~J. 1982, \apj, 252, 
102

\bibitem[]{} 
Condon, J.~J., Huang, Z.-P., Yin, Q.~F., \& Thuan, T.~X. 1991, \apj, 378, 65

\bibitem[]{} 
Cruz-Gonz\'ales, I., Carrasco, L., Serrano, A., Guichard, J.,
Dultzin-Hacyan, D., \& Bisiacchi, G.~F. 1994, \apjs, 94, 47

\bibitem[]{} 
Dahari, O., \& De Robertis, M.~M. 1988, \apjs, 67, 249

\bibitem[]{} 
de Bruyn, A.~G., \& Wilson, A.~S. 1978, \aa, 64, 433

\bibitem[]{} 
Della Ceca, R., Palumbo, G.~G.~C., Persic, M., Boldt, E.~A., Marshall,
E.~E., \& De Zotti, G. 1990, \apjs, 72, 471

\bibitem[]{} 
de Vries, W.~H., O'Dea, C.~P., Barthel, P.~D., Fanti, C., Fanti, R., \&
Lehnert, M.~D. 2000, \aj, 120, 2300

\bibitem[]{} 
Edelson, R. 1987, \apj, 313, 651

\bibitem[]{} 
Elvis, M., Wilkes, B.~J., McDowell, J.~C., Green, R.~F., Bechtold, J., 
Willner, S.~P., Oey, M.~S., Polomski, E., \& Cutri, R. 1994, \apjs, 95, 1

\bibitem[]{} 
Falcke, H., \& Biermann, P.~L. 1995, \aa, 293, 665

\bibitem[]{} 
Feigelson, E.~D., \& Berg, C.~J. 1983, \apj, 269, 400

\bibitem[]{} 
Feigelson, E.~D., \& Nelson, P.~I. 1985, \apj, 293, 192

\bibitem[]{}
Filho, M.~E., Barthel, P.~D., \& Ho, L.~C. 2000, \apjs, 129, 93

\bibitem[]{}
Filippenko, A.~V., \& Sargent, W.~L.~W. 1989, \apj, 342, L11

\bibitem[]{}
Gaskell, C.~M., \& Ferland, G.~J. 1984, \pasp, 96, 393

\bibitem[]{}
Giuricin, G., Fadda, D., \& Mezzetti, M. 1996, \apj, 468, 475

\bibitem[]{}
Giuricin, G., Mardirossian, F., Mezzetti, M., \& Bertotti, G. 1990, \apjs,
72, 551

\bibitem[]{}
Goldschmidt, P., Kukula, M.~J., Miller, L., \& Dunlop, J.~S. 1999, \apj,
511, 612

\bibitem[]{}
Goodrich, R.~W., \& Osterbrock, D.~E. 1983, \apj, 269, 416

\bibitem[]{}
Green, R.~F., Schmidt, M., \& Liebert, J. 1986, \apjs, 61, 305

\bibitem[]{}
Halpern, J.~P., \& Steiner, J.~E. 1983, \apj, 269, L37

\bibitem[]{} 
Hamilton, T.~S., Casertano, S., \& Turnshek, D.~A. 2001, \apj, in press

\bibitem[]{} 
Hamuy, M., \& Maza, J. 1987, \aas, 68, 383

\bibitem[]{} 
Ho, L.~C. 1999a,  \apj, 510, 631

\bibitem[]{} 
------. 1999b,  \apj, 516, 672

\bibitem[]{} 
Ho, L.~C., et al. 2001a, \apj, in press 

\bibitem[]{} 
Ho, L.~C., Filippenko, A.~V., \& Sargent, W.~L.~W. 1995, \apjs, 98, 477

\bibitem[]{} 
------. 1996, \apj, 462, 183

\bibitem[]{} 
------. 1997a, \apjs, 112, 315
 
\bibitem[]{} 
------. 1997b, \apj, 487, 568

\bibitem[]{} 
Ho, L.~C., Filippenko, A.~V., Sargent, W.~L.~W., \& Peng, C.~Y. 1997c, \apjs,
112, 391

 
\bibitem[]{} 
Ho, L.~C., Rudnick, G., Rix, H.-W., Shields, J.~C., McIntosh, D.~H.,
Filippenko, A.~V., Sargent, W.~L.~W., \& Eracleous, M. 2000c, \apj, 541, 120

\bibitem[]{} 
Ho, L.~C., \& Ulvestad, J.~S. 2001, \apjs, in press

\bibitem[]{} 
Ho, L.~C., Van Dyk, S.~D., Pooley, G.~G., Sramek, R.~A., \& Weiler, K.~W.
1999, \aj, 118, 843

\bibitem[]{} 
Holtzman, J., \etal 1995, \pasp, 107, 1065

\bibitem[]{} 
Hooper, E.~J., Impey, C.~D., Foltz, C.~B., \& Hewett, P.~C. 1995, \apj, 445, 62

\bibitem[]{} 
Huchra, J.~P., \& Burg, R. 1992, \apj, 393, 90

\bibitem[]{} 
Huchra, J.~P., Davis, M., Latham, D., \& Tonry, J. 1983, \apjs, 52, 89

\bibitem[]{} 
Hummel, E. 1981, \aa, 93, 93

\bibitem[]{} 
Isobe, T., Feigelson, E.~D., \& Nelson, P.~I. 1986, \apj, 306, 490

\bibitem[]{} 
Kellermann, K.~I., Sramek, R.~A., Schmidt, M., Green, R.~F., Shaffer, D.~B.
1994, \aj, 108, 1163

\bibitem[]{} 
Kellermann, K.~I., Sramek, R.~A., Schmidt, M., Shaffer, D.~B., \& Green,
R.~F. 1989, \aj, 98, 1195

\bibitem[]{} 
Krolik, J.~H. 1998, Active Galactic Nuclei (Princeton: Princeton Univ.  Press)

\bibitem[]{} 
Kukula, M.~J., Dunlop, J.~S., Hughes, D.~H., \& Rawlings, S. 1998, \mnras,
297, 366

\bibitem[]{} 
Kukula, M.~J., Pedlar, A., Baum, S.~A., O'Dea, C.~P. 1995, \mnras, 276, 1262

\bibitem[]{} 
Laor, A. 2000, \apj, 543, L111

\bibitem[]{} 
Lawrence, A. 1987, \pasp, 99, 309

\bibitem[]{} 
Lonsdale, C.~J., Smith, H.~E., \& Lonsdale, C.~J. 1995, \apj, 438, 632

\bibitem[]{} 
MacAlpine, G.~M. 1985, in Astrophysics of Active Galaxies and Quasi-Stellar
Objects, ed. J.~S. Miller (Mill Valley, CA: Univ. Science Books), 259

\bibitem[]{} 
MacKenty, J.~W. 1990, \apjs, 72, 231

\bibitem[]{} 
Malkan, M.~A., Gorjian, V., \& Tam, R. 1998, \apjs, 117, 25

\bibitem[]{} 
Matthews, T.~A., Morgan, W.~W., \& Schmidt, M. 1964, \apj, 140, 35

\bibitem[]{} 
McCarthy, P.~J. 1993, \annrev, 31, 639

\bibitem[]{} 
McLure, R.~J., Dunlop, J.~S., Kukula, M.~J., Baum, S.~A., O'Dea, C.~P.,
\& Hughes, D.~H. 1999, \mnras, 308, 377

\bibitem[]{} 
Meurs, E.~J.~A., \& Wilson, A.~S. 1984, \aa, 136, 206

\bibitem[]{} 
Miller, L., Peacock, J.~A., \& Mead, A.~R.~G. 1990, \mnras, 244, 207

\bibitem[]{} 
Miller, P., Rawlings, S., \& Saunders, R. 1993, \mnras, 263, 425

\bibitem[]{} 
Miller, P., Rawlings, S., Saunders, R., \& Eales, S. 1992, \mnras, 254, 93

\bibitem[]{} 
Morris, S.~L., \& Ward, M.~J. 1988, \mnras, 230, 639

\bibitem[]{} 
Mushotzky, R.~F., \& Wandel, A. 1989, \apj, 339, 674

\bibitem[]{} 
Nandra, K., George, I.~M., Mushotzky, R.~F., Turner, T.~J., \& Yaqoob, T.
1997, \apj, 477, 602

\bibitem[]{} 
Nelson, C.~H., MacKenty, J.~W., Simkin, S.~M., \& Griffiths, R.~E. 1996,
\apj, 466, 713

\bibitem[]{} 
Niklas, S., Klein, U., \& Wielebinski, R. 1995, \aa, 293, 56

\bibitem[]{} 
Osterbrock, D.~E. 1981, \apj, 249, 462

\bibitem[]{} 
------. 1984, QJRAS, 25, 1

\bibitem[]{} 
Osterbrock, D.~E., \& Martel, A. 1993, \apj, 414, 552


\bibitem[]{} 
Peterson, B.~M. 1997, An Introduction to Active Galactic Nuclei (Cambridge:
Cambridge Univ. Press)

\bibitem[]{} 
Phillips, A.~C., Illingworth, G.~D., MacKenty, J.~W., \& Franx, M. 1996, \aj,
111, 1566

\bibitem[]{} 
Ravindranath, S., Ho, L.~C., Peng, C.~Y., Filippenko, A.~V., \&
Sargent, W.~L.~W. 2001, \aj, submitted

\bibitem[]{} 
Rawlings, S., \& Saunders, R. 1991, \nat, 349, 138
                          
\bibitem[]{} 
Rawlings, S., Saunders, R., Eales, S.~A., \& Mackay, C.~D. 1989, \mnras,
240, 701

\bibitem[]{} 
Rush, B., Malkan, M.~A., \& Edelson, R.~A. 1996, \apj, 473, 130

\bibitem[]{}
Sadler, E.~M., Jenkins, C.~R., \& Kotanyi, C.~G. 1989, \mnras, 240, 591

\bibitem[]{} 
Schlegel, D.~J., Finkbeiner, D.~P., \& Davis, M. 1998, \apj, 500, 525

\bibitem[]{} 
Schmidt, M., \& Green, R.~F. 1983, \apj, 269, 352

\bibitem[]{} 
Schmitt, J.~H.~M.~M. 1985, \apj, 293, 178

\bibitem[]{} 
Serjeant, S., Rawlings, S., Maddox, S.~J., Baker, J.~C., Clements, D.,
Lacy, M., \& Lilje, P.~B. 1998, \mnras, 294, 494

\bibitem[]{} 
Shuder, J.~M. 1981, \apj, 244, 12

\bibitem[]{} 
Sramek, R.~A., \& Weedman, D.~W. 1980, \apj, 238, 435

\bibitem[]{} 
Stirpe, G.~M., \etal 1994, \aa, 285, 857

\bibitem[]{} 
Stocke, J.~T., Morris, S.~L., Weymann, R.~J., \& Foltz, C.~B. 1992, \apj,
396, 487

\bibitem[]{} 
Strittmatter, P.~A., Hill, P., Pauliny-Toth, I.~I.~K., Steppe, H., \&
Witzel, A. 1980, \aa, 88, L12

\bibitem[]{} 
Tadhunter, C.~N., Morganti, R., Robinson, A., Dickson, R., Villar-Mart\'\i n,
M., \& Fosbury, R.~A.~E. 1998, \mnras, 298, 1035

\bibitem[]{} 
Terashima, Y., Ho, L.~C., \& Ptak, A.~F. 2000, \apj, 539, 161

\bibitem[]{} 
Thean, A., Pedlar, A., Kukula, M.~J., Baum, S.~A., \& O'Dea, C.~P. 2000,
\mnras, 314, 573

\bibitem[]{} 
Tully, R.~B. 1988, Nearby Galaxies Catalog (Cambridge: Cambridge Univ. Press)

\bibitem[]{} 
Ulvestad, J.~S. 1986, \apj, 310, 136

\bibitem[]{} 
Ulvestad, J.~S., Roy, A.~L., Colbert, E.~J.~M., \& Wilson, A.~S. 1998, \apj,
496, 196

\bibitem[]{} 
Ulvestad, J.~S., \& Wilson, A.~S. 1984a, \apj, 278, 544
 
\bibitem[]{} 
------. 1984b, \apj, 285, 439

\bibitem[]{} 
------. 1989, \apj, 343, 659

\bibitem[]{} 
Ulvestad, J.~S., Wrobel, J.~M., Roy, A.~L., Wilson, A.~S., Falcke, H.,
\& Krichbaum, T.~P. 1999, \apj, 517, L81

\bibitem[]{} 
Visnovsky, K.~L., Impey, C.~D., Foltz, C.~B., Hewett, P.~C., Weymann,
R.~J., \& Morris, S.~L. 1992, \apj, 391, 560

\bibitem[]{} 
Wadadekar, Y., \& Kembhavi, A. 1999, \aj, 118, 1435

\bibitem[]{} 
Ward, M., Elvis, M., Fabbiano, G., Carleton, N.~P., Willner, S.~P., \&
Lawrence, A. 1987, \apj, 315, 74

\bibitem[]{} 
Weedman, D.~W. 1976, \apj, 208, 30

\bibitem[]{} 
White, R.~L., \etal 2000, \apjs, 126, 133

\bibitem[]{}
Whittle, M. 1985, \mnras, 213, 33

\bibitem[]{}
------. 1992a, \apjs, 79, 49

\bibitem[]{}
------. 1992b, \apj, 387, 109

\bibitem[]{}
Wilson, A.~S., \&  Willis, A.~J. 1980, \apj, 240, 429

\bibitem[]{}
Winkler, H. 1997, \mnras, 292, 273

\bibitem[]{}
Winkler, H., Glass, I.~S., van Wyk, F., Marang, F., Jones, J.~H.~S.,
Buckley, D.~A.~H., \& Sekiguchi, K. 1992, \mnras, 257, 659

\bibitem[]{}
Wrobel, J.~M. 2000, \apj, 531, 716

\bibitem[]{}
Wrobel, J.~M., \& Heeschen, D.~S. 1991, \aj, 101, 148

\bibitem[]{} 
Xu, C., Livio, M., \& Baum, S.~A. 1999, \aj, 118, 1169

\bibitem[]{} 
Yee, H.~K.~C. 1980, \apj, 241, 894

\bibitem[]{} 
------. 1983, \apj, 272, 473

\bibitem[]{} 
Zirbel, E.~L. 1996, \apj, 473, 713

\end{thebibliography}
\end{document}